\documentclass[aps,prb,reprint,a4paper,citeautoscript,floatfix]{revtex4-2}

\pdfoutput=1						
\usepackage[charter]{mathdesign}	
\usepackage{amsmath}				

\usepackage[pdftex]{graphicx}
\usepackage{microtype}
\usepackage{bm}

\usepackage[svgnames]{xcolor}
\usepackage[
	colorlinks=True,linkcolor=DarkRed,citecolor=ForestGreen,urlcolor=MediumBlue,
	pdfstartview=FitH,bookmarks=False,pdfpagemode=UseNone
]{hyperref}

\begin{document}

\title{
Effects of forward disorder on quasi-1D superconductors
}

\author{Giacomo Morpurgo, Thierry Giamarchi}
\affiliation{
DQMP, University of Geneva, 24 Quai Ernest-Ansermet, CH-1211 Geneva, Switzerland
}

\date{\today}

\begin{abstract}
We study the competition between disorder and singlet superconductivity in a quasi-1D system. We investigate the applicability of the Anderson theorem,  namely that time-reversal conserving (non-magnetic) disorder does not impact the critical temperature, by opposition to time-reversal breaking disorder (magnetic). To do so, we examine a quasi-1d system of spin 1/2 fermions with attractive interactions and forward scattering disorder using field theory (bosonization). By computing the superconducting critical temperature ($T_c$), we find that for non-magnetic disorder, the Anderson theorem also holds in the quasi-1D geometry. In contrast, magnetic disorder has an impact on the critical temperature, which we investigate by deriving renormalization group (RG) equations describing the competition between disorder and interactions. Computing the critical temperature as a function of disorder strength, we observe different regimes depending on the strength of interactions. We discuss possible platforms where this can be observed in cold atoms and condensed matter.
\end{abstract}

\maketitle

\section{Introduction}

Competition between superconductivity and disorder is a very important and fundamental problem. By nature, superconductivity would naively be expected to be robust to disorder. One important question, addressed from the early days of superconductivity \cite{Abrikosov1961,anderson_dirty_superconductor}, is whether the presence of disorder in the normal phase can impede the occurrence of the superconducting phase transition or drastically change its critical temperature. The result, known under the name of Anderson's theorem \cite{anderson_dirty_superconductor} is that for a singlet superconductor obeying the Bardeen-Cooper-Schrieffer (BCS) mechanism, non-magnetic impurities have no effect on the critical temperature, while magnetic impurities decrease the critical temperature, effectively ``destroying'' superconductivity. This behaviour was explained very clearly by Anderson through a symmetry argument. As long as the disorder still respects time-reversal symmetry (as non-magnetic disorder does), it is possible to form a pair of eigenstates that are the time-reversal partners of each other, instead of the usual $k$ and $-k$ states forming a Cooper pair in the pure system. However, as soon as the disorder breaks time-reversal symmetry (as magnetic disorder does), there is no way of forming an equivalent of the Cooper pair, and superconductivity is destroyed. For magnetic disorder, formulas giving the reduction of Tc with the concentration of disorder were given \cite{Abrikosov1961}.  

This dramatic difference in behaviour extends to other types and pairing natures of superconductivity, where both kind of disorder can significantly decrease the critical temperature. The robustness of the superconducting transition to disorder has thus been seen as a probe of the nature and pairing symmetry of the superconducting order parameter in materials as varied as heavy fermions \cite{varma_review_heavy_fermions}, organic superconductors \cite{jerome_comptes_rendus}, high Tc superconductors \cite{allouletal_RMP} and multiband superconductors \cite{Yerin_multiband_superconductivity_disorder}. The combined problem  of disorder and interactions has also been studied theoretically in different settings including the 3D Hubbard model framework (both repulsive and attractive) to compare the competition between different orders in the presence of disorder \cite{huscroft_hubbard_non_mag_disorder,dobrosavljevic_infinited_disorder}, or 2D superconductors \cite{benfatto_XY_bilayers, dubi_nature07}.

However, the situation was realized to be more complicated than predicted by simple applications of the BCS mean-field equation to disordered systems. This is particularly the case when dimensionality allows the disorder to have a strong effect, such as e.g. leading to Anderson localization \cite{anderson_localisation}. In such cases, since the very nature of the eigenstates is affected by the disorder, the symmetry argument does not suffice, and even non-magnetic disorder can potentially affect $T_c$. It was indeed shown to be the case for systems made of coupled 1D chains with attractive interactions. In such a case non-magnetic disorder is able to destroy even s-wave pairing \cite{suzumura_mean_field}, or in some regime even  enhance it \cite{lowe_disorder_enhance_superconductivity_quasi_1d}. 

One could argue that such results are the direct consequence of the existence of the rather drastic phenomenon of Anderson localization, quite efficient in one dimension, but much easier to reduce strongly in higher dimensional cases. Even in one dimension, since Anderson localization is intimately linked to the presence of backscattering due to disorder \cite{berezinskii_conductivity_log,abrikosov_rhyzkin,giamarchi_loc}, one could expect to recover an Anderson theorem if such backscattering is suppressed and if only forward scattering exists from the disorder. 

Such questions have become timely since recently cold atomic systems have provided excellent experimental realizations in which both disorder and interactions could be controlled \cite{Bloch2008,esslinger_annrev_2010}. In such systems, localization of one particle in quasiperiodic or speckle potentials \cite{roati08_anderson_localization_BEC,aspect_anderson_localization_BEC.pdf,sanchez-palencia_review_disorder_cold,Piraud_Anderson_speckle}, disordered interacting bosons \cite{fallani_biperiodic_cold,derrico_cold_boseglass} and fermions \cite{jendrzejewski_3d_anderson_localisazion_cold_atoms,schreiber_manybody_localization_cold} have been realized. More generally, cold atoms have proven to be excellent systems to probe or think of combined effects of interactions and disorder or quasiperiodicity in a large variety of situations \cite{yao_quasiperiodic_1d_equal_potential,yao_quasiperiodic_equal_potential_bose_glass, Sbroscia_2d_quasicrstal_localization}.

In this paper, we thus examine the effects of non-magnetic and magnetic disorder on a system made of coupled one dimensional fermionic chains with attractive interactions. To avoid or minimize effects due to Anderson localisation, we restrict ourselves only to long wavelength disorder having Fourier components much smaller than $2 k_F$, where $k_F$ is the Fermi wavevector of the chains. In Sec.~\ref{sec:Model}, we present the model, the bosonized formalism and the observables that we will look at in order to compute the critical temperature. In Sec.~\ref{sec:non_magnetic_disorder}, we examine non-magnetic disorder and find that forward non-magnetic disorder has no impact on the critical temperature. In Sec.~\ref{sec:magnetic_disorder}, we instead study the case of magnetic disorder. First, we look at the case where we neglect the spin gap in the spin sector of the Hamiltonian, which simplifies the problem and allows us to treat it almost completely analytically. We find that disorder here weakens the superconductivity until it destroys it. Secondly, we treat the case of a finite spin gap. To deal with the corresponding ``sine-Gordon'' term in the Hamiltonian, we use a renormalization group (RG) technique. In this case, while we still find that disorder will ultimately lead to the destruction of superconductivity, we also see that depending on the strength of interactions we get two very distinct regimes arising from the competition of the disorder with the spin gap due to the interactions. In Sec.~\ref{sec:Discussion}, we discuss several aspects of our results and also how they could be practically implemented in a cold atom or condensed matter realization. Finally, a conclusion can be found in Sec.~\ref{sec:conclusion} and more technical details can be found in the appendices. 

\section{Model} \label{sec:Model}
\subsection{General microscopic model}
\label{sec:microscopic_model}

\begin{figure}
\includegraphics[width=\columnwidth]{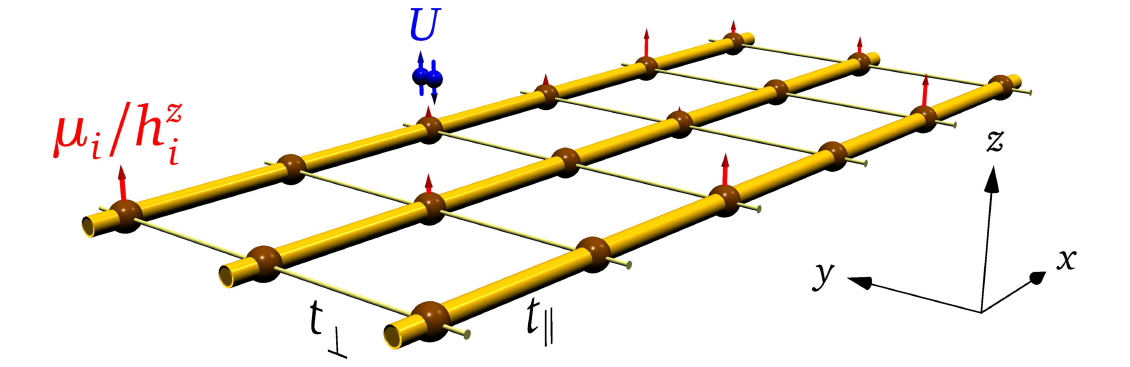}
\caption{\label{fig:model}
Fermionic tubes of spin $1/2$ particles which can tunnel between the tubes. The particles experience a contact attractive interaction $U$ (see text), leading to a singlet superconducting ground state. The tubes can be continuous, or have their own lattice. In which case the model is the attractive Hubbard model with anisotropic hopping $t_\parallel$ along the chains and $t_\perp$ between the chains. We are in a situation where $t_\perp \ll t_\parallel$. Additionally, along the chains, there is either a random chemical potential $\mu_i$ or a random magnetic field $h_i^z$.}
\end{figure}

We consider a fermionic spin 1/2 model with attractive contact interaction $U<0$, made of several 1d tubes arranged in a 3D lattice. The system is schematically shown in Fig.~\ref{fig:model}. 
The ground state of the pure system is thus a singlet superconductor with, in particular, a gap in the spin excitations \cite{giamarchi_book_1d}. The 1D tubes can be continuous or with a lattice, both cases can be treated similarly.

We consider two types of disorder. One is non-magnetic disorder where a random chemical potential couples to the total density 
\begin{equation}
\rho_n(x) = c^{\dagger}_{\uparrow,n}(x)c^{}_{\uparrow,n}(x) + c^{\dagger}_{\downarrow,n}(x)c^{}_{\downarrow,n}(x)
\end{equation}
where $c^\dagger_{\sigma,n}(x)$ creates a fermion with spin $\sigma$ at point $x$ on chain $n$. The corresponding disorder term is
\begin{equation}
 H_{\text{dis}, \rho} = - \sum_n \int dx \, \mu_n(x) \rho_n(x)
\end{equation}
The second type is a random magnetic field coupling to the spin density along $z$
\begin{equation}
\sigma^z_n(x) = c^{\dagger}_{\uparrow,n}(x)c^{}_{\uparrow,n}(x) - c^{\dagger}_{\downarrow,n}(x)c^{}_{\downarrow,n}(x)
\end{equation}
leading to 
\begin{equation}
 H_{\text{dis},\sigma} = - \sum_n \int dx \, h^z_n(x) \sigma^z_n(x)
\end{equation}

Both the chemical potential $\mu_n(x)$ and random magnetic field $h^z_\mu(x)$ are taken to be uncorrelated from chain to chain. For the correlation of the disorder along the chains, we wish to avoid the dominant effects of Anderson localization, which is produced by the backscattering on the disorder, and which is anomalously strong in one dimension. As discussed in Ref.~\onlinecite{suzumura_mean_field,lowe_disorder_enhance_superconductivity_quasi_1d}, this has a drastic effect even on plain vanilla singlet superconductors. We thus restrict the disorder to have only Fourier components much smaller than $2 k_F$, where $k_F$ is Fermi wavevector of one chain. We will discuss the consequences of such a restriction on disorder in more details in Sec.~\ref{sec:implementations}, but we just note here that such a limitation of the spectrum of disorder is quite natural with speckle disorder \cite{lugan_correlated_potentials}.

\subsection{Bosonized representation}
\label{sec:bosonized_model}

To deal with the interactions in each chain, we use the bosonization technique \cite{giamarchi_book_1d} and introduce collective variables $\phi_{\rho,n}$ (resp. $\phi_{\sigma,n}$) linked to the fluctuations of charge (resp. spin) density on chain $n$ by 
\begin{equation} \label{eq:phis}
\begin{split}
 \rho_n(x) &= \frac{-\sqrt2}{\pi} \nabla\phi_{\rho,n}(x) \\
 \sigma_n^z(x) &= \frac{-\sqrt2}{\pi} \nabla\phi_{\sigma,n}(x)
\end{split}
\end{equation}
These variables are conjugate to variables $\theta_{\nu,n}$ (with $\nu=\rho,\sigma$) linked to the phase fluctuations and which obey the canonical commutation relations $[\phi_{\nu,n}(x),\nabla\theta_\nu(x',n)] =i \pi \delta(x-x')$.

After bosonizing each chain individually, we obtain the following Hamiltonian: 
\begin{equation}\label{eq:hamtot}
 H = \sum_{n} H_{1D,n}+ \sum_{\langle n,l\rangle}H_{\perp, nl}
\end{equation}
where $\langle n,l \rangle$ denotes chains that are nearest neighbors on a lattice. 
The Hamiltonian of a single chain is 
\begin{equation}\label{eq:singlechain}
\begin{split}
    H_{1D,n} &= \frac{1}{2\pi}\int dx \,\left[ \frac{u_{\rho}}{K_\rho}\left(\nabla \phi_{\rho,n} \right)^2 + u_{\rho}K_{\rho}\left( \nabla \theta_{\rho,n} \right)^2\right]\\
    &+ \frac{1}{2\pi}\int dx \, \left[\frac{u_{\sigma}}{K_\sigma}\left( \nabla \phi_{\sigma,n} \right)^2 + u_{\sigma}K_{\sigma}\left( \nabla \theta_{\sigma,n} \right)^2 \right]\\
    &+ \frac{2g}{(2\pi\alpha)^2}\int \, dx \cos(\sqrt{8}\phi_{\sigma,n})\\
    & + H_{\text{dis},n, \rho/\sigma}
\end{split}
\end{equation}
The parameters $K_\rho$, $K_\sigma$, $u_\rho$, and $u_\sigma$ are the Luttinger parameters and contain the effects of the interactions and the kinetic energy. $\alpha$ is linked to the ultraviolet cut-off of our bosonic theory.

The forward disorder Hamiltonian reads in bosonized form
\begin{align}
 H_{\text{dis},n,\rho} &=-\frac{\sqrt{2}}{\pi}\int dx \, \eta_n(x) \nabla \phi_{\rho,n} \label{eq:discharge}\\
  H_{\text{dis},n,\sigma} &=-\frac{\sqrt{2}}{\pi}\int dx \, \gamma_{z,n}(x) \nabla \phi_{\sigma,n} \label{eq:disspin}
\end{align}
where $\gamma_z$ and $\eta$ are 2 random Gaussian fields with correlations $\overline{\gamma_{z,n}(x)\gamma_{z,m}(x')}=D_{f,e}\delta(x-x')\delta_{n,m}$, $\overline{\eta_n(x)\eta_m(x')}=D_{f,m}\delta(x-x')\delta_{n,m}$ characterizing the forward magnetic and non-magnetic disorders,

Finally, we have to consider the interchain coupling. The elementary coupling between the chains should be produced by single particle hopping. However, we consider in this paper a different coupling between the chains, namely we retain only the tunnelling of pairs by Josephson coupling. The corresponding part of the Hamiltonian is thus
\begin{equation} \label{eq:interchain}
    H_{\perp,nl} = -J\int dx \, \left[
    c^\dagger_{\uparrow,n}(x)c^\dagger_{\downarrow,n}(x)
    c^{}_{\downarrow,l}(x)c^{}_{\uparrow,l}(x)\right]
\end{equation}
where $n,l$ are the 1D chains indexes, and $J$ the Josephson coupling between the chains. We take here $J$ as an independent parameter. 

We will come back in Sec.~\ref{sec:interjos} on this point. Note that retaining only the Josephson coupling and discarding the single particle tunnelling is usually justified by the presence of a gap in the spin sector ($\Delta_\sigma$) for attractive interactions \cite{giamarchi_book_1d}. We will take (\ref{eq:interchain}) as the interchain Hamiltonian for our study, while still noting that in the case of a magnetic disorder, which can potentially break the spin gap the situation is more delicate. For large negative $U$, one has the usual estimate of the Josephson coupling $J \sim  t_\perp^2/|\Delta_\sigma|$ where $\Delta_\sigma \sim U$. While another intermediate regime where the Josephson coupling has a different dependence on the spin gap ($J \sim  t_\perp^2/|\Delta_\sigma|^2$)  exists \cite{bourbonnais_lettre_tperp_RG}. By taking $J$ constant, we potentially ignore possible effects of having a Josephson coupling reinforced by the disorder if the spin gap decreases.

The physical properties of the system are thus controlled by the above Hamiltonian and thus depend crucially on the Luttinger liquid parameters. In order to work with a specific example, we focus in the following to the parameters corresponding to the case of tubes with a lattice that realizes the fermionic spin-1/2 Hubbard model with the Hamiltonian
\begin{equation}\label{eq:hamdep}
    H=-\sum_{\langle i,j \rangle,\sigma}t_{i,j} c^{\dagger}_{i,\sigma} c^{}_{j,\sigma} + h.c. + U\sum_{i}c^{\dagger}_{i,\uparrow}c^{}_{i,\uparrow}c^{\dagger}_{i,\downarrow}c^{}_{i,\downarrow} + H_{\text{dis}}
\end{equation}
where $t_{i,j}$ is the hopping amplitude from site $i$ to site $j$, $c_{i,\sigma},c^\dagger_{i,\sigma}$ are the destruction/creation operators at site $i$ and spin $\sigma$, $U$ is the on-site interaction, and $\langle i,j \rangle$ denotes nearest neighbors on a cubic lattice. By quasi 1d system, we mean that the hopping $t_{i,j}$ is small in all but one direction $t_z = t_y \ll t_x$. $H_{\text{dis}}$ encodes the disorder part of the Hamiltonian.

For a clean 1D attractive Hubbard model, the spin sector is gapped and the parameters of the charge sector can be computed perturbatively in $U$ \cite{giamarchi_attract_1d,giamarchi_book_1d} and lead to:
\begin{equation}
\label{eq:Hubbard_parameters}
\begin{split}
    K_{\rho,\sigma} &= (1 \pm U/(\pi v_F))^{-1/2}\\
    u_{\rho,\sigma} &= v_F(1 \pm U/(\pi v_F))^{1/2}\\
    g & = U
\end{split}
\end{equation}
where the upper sign is for $\rho$ (the lower sign is for $\sigma$) and $v_F$ is the Fermi velocity $v_F = 2 t_\parallel \sin(k_F)$. 
 
\subsection{Observables}

Since our main goal is to compute the effect of the disorder on the superconducting critical temperature, a central part of our calculations is the pair correlation function. 

We treat the interchain Hamiltonian (\ref{eq:interchain}) in mean-field 
\begin{equation}
 c^\dagger_{\uparrow,n}(x)c^{\dagger}_{\downarrow,n}(x) = \langle c^\dagger_{\uparrow,n}(x)c^{\dagger}_{\downarrow,n}(x) \rangle + \delta \left(c^\dagger_{\uparrow,n}(x) c^{\dagger}_{\downarrow,n}(x) \right)
\end{equation}
and retain only the terms linear in the fluctuation part around the mean value. Note that here, the order parameter $\langle c^\dagger_{\uparrow,n}(x)c^{\dagger}_{\downarrow,n}(x) \rangle$ is dependent on space for a single realisation over the disorder. However, once we do the average over the different disorder realizations, since the disorder is decoupled between the different chains, we recover an order parameter which will be space invariant.

The mean field approximation leads to 
\begin{equation}
    H_{\perp,MF} = -\frac{2zJ}{\pi\alpha} \Delta \sum_n  \cos\left(\sqrt{2}\theta_{\rho,n}(x)\right)\cos\left(\sqrt{2}\phi_{\sigma,n}(x)\right)
\end{equation}
where $z$ is the number of neighboring chains and 

\begin{equation}
    \Delta = \frac{1}{\pi\alpha} \langle e^{i\sqrt{2}\theta_\rho}\cos\left(\sqrt{2}\phi_\sigma\right)\rangle
\end{equation}
Where we will choose the gauge where $\Delta$ is real.

The critical temperature is given by the divergence of the pair susceptibility $\chi(T)$, which is given in the mean-field (RPA) approximation by 
\begin{equation}
    \chi(\beta) = \frac{\chi_0(\beta)}{1-\frac{2zJ}{(\pi\alpha)^2}\chi_0(\beta)}
\end{equation}
where $\chi_0(\beta)$ is the uniform and static susceptibility in the absence of interchain coupling at the temperature $T=1/\beta$. The superconducting critical temperature $T_c$ is thus given by the condition 
\begin{equation} \label{eq:meandieldeq}
    1 = \frac{2zJ}{(\pi\alpha)^2}\chi_0(\beta_c)
\end{equation}
where 
\begin{multline}
\chi_0(\beta) = \int dx \int^{\beta}_0 d\tau \\
    \langle T_\tau \cos\left(\sqrt{2}\theta_\rho(x,\tau)\right)\cos\left(\sqrt{2}\phi_\sigma(x,\tau)\right) \\
    \cos\left(\sqrt{2}\theta_\rho(0,0)\right)\cos\left(\sqrt{2}\phi_{\sigma}(0,0)\right)\rangle_{H_{1D}}
\end{multline}
Since the Hamiltonian (\ref{eq:singlechain}) is separated between the charge and spin sectors we obtain 
\begin{multline} \label{eq:correlsep}
 \chi_0(\beta) = \frac{1}{u_\sigma}\int\int_{\Gamma(\beta)} dx d(u_\sigma\tau) \\
    \langle T_\tau \cos\left(\sqrt{2}\theta_\rho(x,u_\rho\tau)\right) \cos\left(\sqrt{2}\theta_\rho(0,0)\right)\rangle_{H_{\rho}}\\
    \cdot\langle T_\tau \cos\left(\sqrt{2}\phi_\sigma(x,u_\sigma\tau)\right) \cos\left(\sqrt{2}\phi_\sigma(0,0)\right)\rangle_{H_\sigma}
\end{multline}
where we have used the fact that for distances which are larger than $u_\sigma\beta$, the correlations decay exponentially in 1D. Therefore we can neglect this part and integrate on a disk of radius $u_\sigma\beta$ denoted $\Gamma(\beta)$ .

Note that this mean-field solution is linked to the quasi-one dimensional nature of the problem and is different from the usual BCS mean-field calculation that was used to establish the Anderson theorem \cite{anderson_dirty_superconductor,Abrikosov1961}.
In the latter case one computes the pair susceptibility with the \emph{full} kinetic energy (thus including the transverse hopping as well) but in an RPA approximation in the \emph{interaction} $U$. At $T_c$, the gap is thus automatically zero in the BCS approximation. In the quasi-one dimensional situation we consider here, the situation is different since even for decoupled chains, for which the $T_c$ is zero even in the absence of disorder, due to the quantum fluctuations, a strong spin gap can exist for the spin sector. We discuss more these differences in the Sec.~\ref{sec:BCS_differences}. 

\section{Non-magnetic disorder}
\label{sec:non_magnetic_disorder}

Let us first consider the case of non-magnetic disorder. The Hamiltonian to consider at the single chain level, in particular to solve (\ref{eq:meandieldeq}) giving $T_c$, is (\ref{eq:singlechain}) with the disorder (\ref{eq:discharge}). 

Since the charge sector part of the Hamiltonian is purely quadratic, the disorder can be absorbed by a simple redefinition of the field $\phi(x)$ \cite{giamarchi_loc}
\begin{equation} \label{eq:absorbcharge}
    \phi_\rho \to \tilde{\phi_\rho}=\phi_\rho-\frac{\sqrt{2}K_\rho}{u_\rho}\int_0^x dx'\eta(x')
\end{equation}
This leads to an Hamiltonian for the charge sector of the form:
\begin{equation} \label{eq:Hamilton_dis_charge}
\begin{split}
    H &= \frac{1}{2\pi}\int dx \, \frac{u_{\rho}}{K_\rho}\left( \nabla \tilde{\phi}_\rho \right)^2 + u_{\rho}K_{\rho}\left( \nabla \theta_{\rho} \right)^2\\
    & -\frac{K_{\rho}}{u_{\rho}\pi}\int dx \, \eta^2(x)\\
\end{split}
\end{equation}

Note that this transformation does not affect the field $\theta(x)$, which remains conjugate to the field $\tilde\phi(x)$. The calculations of the susceptibility, which depends on $\langle T_\tau \cos(\sqrt{2}\theta_\rho(x,u\tau)) \cos(\sqrt{2}\theta_\rho(0,0))\rangle_{H_{\rho}}$, and the spin part are thus not affected  by the disorder either, and yield an identical result with or without forward disorder. 

An analogous result to the Anderson theorem, namely that a non-magnetic forward disorder has no impact on the critical temperature of superconductivity ($T_c(D_{f,e})/T_c(0) = 1$) is recovered. This result can also be viewed directly in the fermion language since the transformation (\ref{eq:absorbcharge}) correspond to a redefinition 
\begin{equation} \label{eq:gaugefermions}
    \psi_{R,L}(r) \to \psi_{R,L} e^{\pm i \frac{K_\rho}{u_\rho}\int_{-\infty}^x dy \eta(y)}
\end{equation}
where $R$ (resp. $L$) denotes the right and left movers and in (\ref{eq:gaugefermions}) the upper sign refers to $R$. Similarly to the Anderson theorem, one thus see that it is possible to create new objects that are still related by time reveral symmetry, even in the presence of disorder. When pairing these objects, the disorder totally disappears, leading to the invariance of the critical temperature. However, the forward scattering disorder will still affect other correlations in this model, for example the density-density ones (basically anything which involves the field $\phi_\sigma$).

Note that this result is modified if the backward scattering is present \cite{suzumura_mean_field}. The Anderson localization that it induces leads to an exponential decay of the pair correlation functions and thus compete with the superconductivity. Therefore, one can expect drastically different effects of non-magnetic disorder on the superconductivity depending on which Fourier components are present. This can, in principle, be tested by changing $k_F$ with respect to an upper cutoff in the disorder spectrum. 

\section{Magnetic disorder}
\label{sec:magnetic_disorder}

We now turn to the case of magnetic disorder. For this, we employ (\ref{eq:singlechain}) with the disorder (\ref{eq:disspin}). Two significant differences are immediately noticeable compared to the case of non-magnetic disorder. Firstly, the spin sector of the Hamiltonian (\ref{eq:singlechain}) is not simply quadratic but has a sine-Gordon form. Thus, an analogous transformation to (\ref{eq:absorbcharge}) done for the spin sector does not allow to get rid of the magnetic disorder in the Hamiltonian. This reflects the competition between the random magnetic field and the cosine term that creates the spin gap. A corresponding term would exist in the charge sector only if the system is in a Mott state with a commensurate filling \cite{orignac_mg_short,giamarchi_mottglass_long}. 

Secondly, the pair susceptibility (\ref{eq:correlsep}) dependson the field $\phi_\sigma$ for the spin sector. This contrasts with the charge sector where the dual field $\theta_\rho$ appears. Thus, performing the afore mentioned transformation introduces a disorder dependence in the pair susceptibility, regardless of the presence of the $\cos(\sqrt8\phi_\sigma)$ term in the Hamiltonian. This indicates from the start that magnetic disorder affects the correlations and consequently the critical temperature. 

In the calculation of the pair susceptibility, the charge sector Hamiltonian is quadratic. Therefore, the charge part of the correlations is \cite{giamarchi_book_1d}
\begin{multline}
    R_\theta(r)=\langle T_\tau \cos\left(\sqrt{2}\theta_\rho(x,u\tau)\right) \cos\left(\sqrt{2}\theta_\rho(0,0)\right)\rangle_{H_{\rho}} \\
    \simeq \frac{1}{2}\left(\frac{\alpha}{r}\right)^{\frac{1}{K_\rho}}
\end{multline}
where $r$ is given by $(x,u_{\rho/\sigma}\tau)$ depending on where we are computing correlation functions (charge or spin sector) and $r^2=x^2 + (u_{\rho/\sigma}\tau)^2$.

Due to the sine-Gordon form of the spin part of the Hamiltonian (\ref{eq:singlechain}), the full calculation of the spin sector is more involved, and we analyze it in the next two sections. 

\subsection{Spin sector with $g=0$}\label{sec:g=0}

Let us start in this section by setting $g=0$ in the Hamiltonian. This amounts to a scenario where the spin gap opened by the presence of the cosine term is so small that it can be neglected. This corresponds typically to the case of attractive interactions very small compared to the kinetic energy in the chain since, in such cases, it is well known that the spin gap is exponentially small in the ratio $t_\parallel/|U|$. This simplified model allows to disentangle the effects produced by the magnetic disorder on the pair susceptibility from the robustness of the finite pairing gap. 

It is important to note that setting $g=0$ does not necessarily imply that $T_c$ itself is small, since $Tc$ is given by equation (\ref{eq:meandieldeq}) and $\chi_o$ depends mostly on $K_\rho$ and $K_\sigma$. Although in this case we can formally take any value for $K_\sigma$, we restrict ourselves to $K_\sigma = 1$, which corresponds to a spin rotation invariant Hamiltonian with $g\to 0$ \cite{giamarchi_book_1d}. Our study can be readily extended to any value of $K_\sigma$ when $g=0$. 

Furthermore, as mentioned in the previous section, we still consider, even if we take a zero spin gap, that the interchain coupling is of the Josephson form, with a fixed $J$. We come back on this approximation in Sec.\ref{sec:interjos}.

The dephasing induced by the disorder (\ref{eq:disspin}) can be readily computed in the case of $g=0$. It affects  the spin part of the correlations and consequently Tc. For $g = 0$, a change of variables similar to the one performed in the charge sector is made :
\begin{equation} \label{eq:shiftspin}
    \phi_\sigma \to \tilde{\phi_\sigma}=\phi_\sigma-\frac{\sqrt{2}}{u_\sigma}\int_0^x dx'\gamma_z(x')
\end{equation}
The transformation removes the disorder from the Hamiltonian. Unlike the case of non-magnetic disorder, the change of variables (\ref{eq:shiftspin}) modifies the susceptibility (\ref{eq:correlsep}). After performing the ensemble averaging over disorder, these correlations become :
\begin{multline}
\label{eq:correlation_spin_g_0}
\langle T_\tau \cos\left(\sqrt{2}\phi_\sigma(x,u\tau)\right) \cos\left(\sqrt{2}\phi_\sigma(0,0)\right)\rangle_{H_{\sigma}} \\ \simeq \frac{1}{2}\left(\frac{\alpha}{r}\right)e^{-\frac{2D_{f,m}}{u_\sigma^2}|x|}
\end{multline}
The forward magnetic disorder thus leads to an exponential decay of the spin part of the correlation function with a characteristic lengthscale related to the disorder strength. Consequently, one can expect a strong impact of magnetic impurities on $T_c$, roughly resulting in a suppression of the superconducting critical temperature when the thermal length associated with the temperature $T_c$ becomes larger that the length associated with the disorder. 

More quantitatively, to determine $T_c$ we solve (\ref{eq:meandieldeq}), which with $g=0$ is :
\begin{equation} \label{eq:selfgzero}
\begin{split}
    1 &= \frac{2 z J}{(\pi \alpha)^2} \int dx \int_0^{\beta_c} d\tau \frac{1}{4}\left(\frac{\alpha}{r}\right)^{\frac{1}{K_\rho} + 1} e^{-\frac{2D_{f,m}}{u_\sigma^2}|x|} \\
     &= \frac{zJ}{u_\sigma(2\pi \alpha)^2}\alpha^{1+\frac{1}{K_\rho}} \int_\alpha^{u/T_c}dr r^{-\frac{1}{K_\rho}}2\pi 
     F\left(\frac{2D_{f,m}}{u_\sigma^2}r\right)
\end{split}
\end{equation}
where $F(x) = (I_o(x)-L_o(x))$ and $I_o$ (resp. $L_o$) is a modified Bessel function of the 1st kind (resp. a modified Struve function). For simplicity we have assumed in the above formula that the charge velocity $u_\rho$ and the spin one $u_\sigma$ are identical so we can use the same $r$ for the spin and charge sector. These two velocities are in general different \cite{giamarchi_book_1d}. The generalization of (\ref{eq:selfgzero}) to the case of two different velocities is straightforward but doesn't change fundamentally the results, albeit at the cost of not having analytically closed expressions, We discuss it in the Appendix~\ref{sec:app_diff_velocities}.

To compute numerically the integrals (\ref{eq:selfgzero}), one has to be especially cautious. Indeed, while the difference of Bessel and Struve functions is well behaved, these two functions individually diverge exponentially. One has to implement a series expansion of the difference at large argument to compute the integrals accurately. To achieve a good numerical convergence we first perform the integral on $r$ and only in a second time the integral on the angle $\theta$, using polar coordinates for 
(\ref{eq:selfgzero}).

It is also worth noting that although the disorder leads to an exponential decay in space, its time independence preserves a good temporal coherence, leading ultimately after integration over the polar angle to  a power-law decay of the correlation at large distances $\lim_{x\to \infty}  F(x) = \frac{2}{\pi x}$. The decay induced by the disorder is thus less dramatic than could have been expected.

To understand the results, let us introduce several characteristic lengths. On one hand $\xi_D = \frac{u_\sigma^2}{2D_{f,m}}$ is the disorder length, which controls the exponential decay rate in (\ref{eq:correlation_spin_g_0}). On the other hand, $\xi_J = \frac{v_F}{T_c(0)}$ is a length associated to the superconducting state in the absence of  disorder. We also define $D_c$ as the critical disorder, the amount of disorder where there is no finite $T_c$ associated with the superconducting phase transition.

\begin{figure}
\includegraphics[width=\columnwidth]{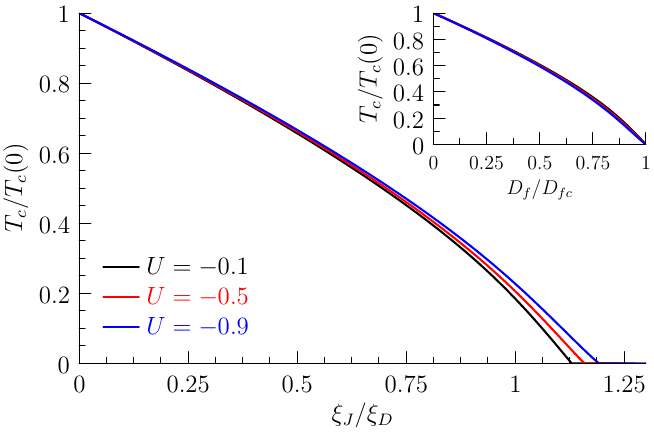}
\caption{\label{fig:Tc_g=0}
$T_c/T_c(0)$ as a function of the ratio of characteristic lengths for different interactions $U$. The inset shows  the same quantity as a function of the disorder normalized by the critical value of the disorder $D_c$ at which superconductivity is destroyed.}
\end{figure}

Fig.~\ref{fig:Tc_g=0} shows the ratio $T_c/T_c(0)$, the critical temperature normalized by the one without disorder, as a function of $\xi_J/\xi_D$ for different values of the attractive interaction $U$, showing the destruction of the superconductive state by the magnetic disorder. The curves depend on the interaction, with the more attractive cases being slightly more robust to disorder. The inset shows the same effect as a function of $D/D_c$. While the curves do not perfectly collapse on each other, they show an excellent 
scaling in these variables. 

The limit of $g=0$ is thus a simple limit showing clearly the analogy for the quasi-1D situation of the Anderson theorem. Although the non-magnetic (forward) disorder has no effect on $T_c$, the magnetic one (time reversal breaking) impacts the critical temperature. Note that these effects are not connected to Anderson localization since they are produced by forward scattering on the disorder.  

\subsection{The general case $g\neq 0$}
The calculation of the previous section actually overestimates the effects of magnetic disorder, because the full Hamiltonian with a \emph{finite} attractive interaction creates a spin gap via the $\cos(\sqrt{8}\phi_\sigma)$ term. This spin gap, which locks the particles in singlet states, prevents the magnetic field from acting. Note that in the quasi-1D geometry that we consider, the critical temperature is controlled by the interchain Josephson coupling, while the spin gap essentially depends on the ratio $t_\parallel/|U|$. Thus, it is perfectly possible to have a large spin gap and a small $T_c$, contrarily to what happens if one computes the $T_c$ in the BCS approximation, for which the spin gap is essentially zero close to $T_c$. This situation is very similar to the case of the attractive higher dimensional Hubbard model, for which a large regime of pseudogap can exist above $T_c$ when $|U|$ is large. 

Determining the effect on $T_c$ is more involved in the case $g\neq 0$. We describe in this section a renormalization group method allowing the calculation of the correlations in the spin sector. 

To renormalize the susceptibility $\chi_o$, we follow a similar procedure than the one used in \cite{giamarchi_logs}. The correlation function $R_\sigma$ is given for $g=0$ by 
\begin{equation}
    R_\sigma(r) = \frac{1}{2}e^{-\frac{K_\sigma}{2}\ln\left(\frac{x^2+(u_\sigma|\tau|+\alpha)^2}{\alpha^2}\right)}e^{\frac{-2K_\sigma^2D_{f,m}}{u_\sigma^2}|x|}
\end{equation}
For $r \gg \alpha$, we have $x^2+(u_\sigma|\tau|+\alpha)^2 \approx r^2$. We consider the function 
\begin{equation} \label{eq:basefunc}
H_\sigma(r) =  R_\sigma(r)e^{\frac{2K_\sigma^2D_{f,m}}{u_\sigma^2}|x|}e^{\frac{K_\sigma}{2}\ln\left(\frac{x^2+(u_\sigma|\tau|+\alpha)^2}{\alpha^2}\right)}
\end{equation} 
with $g=0$, $H_\sigma =1/2$. For $g\neq0$, we renormalize the cutoff $\alpha$ until it reaches $r$ in the perturbative expansion of (\ref{eq:basefunc}) in powers of $g$. This multiplicative renormalization procedure allows to define a function $I_\sigma(dl,g(l))$ such that 
\begin{equation}
    H_\sigma(r,\alpha_o,g(\alpha_o)) = e^{\int_0^{\ln(r/\alpha)}\ln(I(dl,g(l)))}
\end{equation}

\subsubsection{RG flow}
The algebra can be found in Appendix~\ref{app:RG_equations}.

The renormalization equations for the parameters $K_\sigma$, $g$, $D_{f,m}$ read:
\begin{align}
     \frac{dK_\sigma}{dl} &= -\frac{g^2K_\sigma^2}{2\pi^2u_\sigma^2}F\left(\frac{8D_{f,m}K_\sigma^2\alpha_o}{u_\sigma^2}e^l\right)\label{eq:RG_K}\\
     \frac{dg}{dl} &= (2-2K_\sigma)g\label{eq:RG_g}\\
    \frac{dD_{f,m}}{dl} &= -\frac{g^2K_\sigma D_{f,m}}{\pi^3u_\sigma^2}G\left(\frac{8D_{f,m}K_\sigma^2\alpha_o}{u_\sigma^2}e^l\right) \label{eq:RG_disorder}\\
    F(x)&=\left(I_o\left(x\right)-L_o\left(x\right)\right) \label{eq:definition_F}\\
    G(x)&=-\frac{2}{3}x+\pi\frac{I_1(x)-L_1(x)}{x}+\pi\left(I_2(x)-L_2(x)\right)\label{eq:definition_G}
\end{align}
where $I_o, I_1$ and $I_2$ are modified Bessel functions of the 1st kind, and $L_o, L_1$ and $L_2$ are modified Struve functions.

Note that for simplicity, we neglect the renormalization of the $u_\sigma$ parameter in our calculation. Indeed we don't expect the renormalization of the speeds to be large, and we also don't expect it to lead to new physical phenomena.

In the limit of no-disorder $D_{f,m}\to 0$, we recover the usual RG equations for $K_\sigma, g$ because $F(0)=1$. Here, we directly see that the disorder and the $g$ term are competing against each other. The $g$ term controls the decrease of the disorder (\ref{eq:RG_disorder}). On the other hand, the effect of the disorder on $g$ is more subtle. The RG equation for $g$ (\ref{eq:RG_g}) depends only on $K_\sigma$, whereas if $K_\sigma$ is smaller than $1$, $g$ is relevant, and irrelevant otherwise. However, the disorder acts on the RG equation for $K_\sigma$  (\ref{eq:RG_K}) and slows down its decrease ($F(x)$ is always smaller than 1)~! It therefore indirectly opposes itself to the parameter reaching a regime where $g$ would be relevant (or at least slow down the $g$ divergence if $g$ is relevant). More details on the interpretation of the RG equations are given in the Appendix~\ref{sec:app_interpretation_RG_equation}.

\begin{figure}
 \includegraphics[width=\columnwidth]{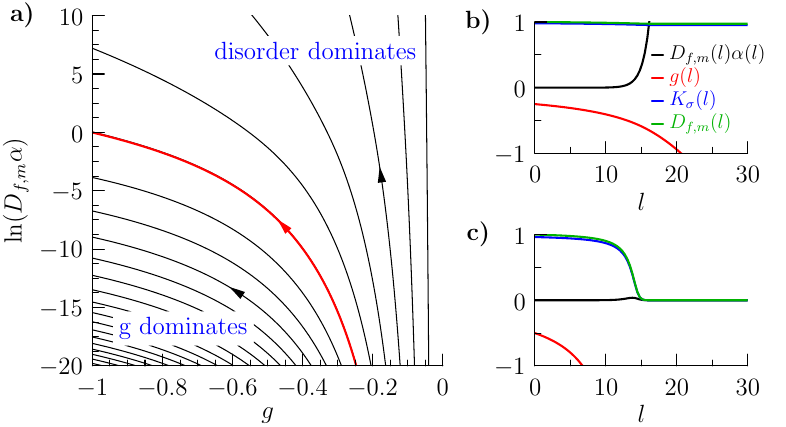}
 \caption{\label{fig:RG_flow}
\textbf{a)} RG flow of our model, in red, the separatrix between the two different regions (disorder dominating vs $g$ dominating); \textbf{b)} Example of the flow for parameters for which the disorder dominates; \textbf{c)} Example of the flow in the $g$ dominated regime.}
\end{figure}

Fig.~\ref{fig:RG_flow}.a shows the RG flow in the $D_{f,m} \alpha-g$ plane. We separate it in two zones, delimited by a separatrix (red line), defined by which one of those quantities (in absolute values) reaches 1 first. 

Below the separatrix, as seen in Fig.~\ref{fig:RG_flow}.c, the term in $g$ diverges first. This can be interpreted as the system succeeding in having a spin gap due to the attractive interaction, with the disorder being too weak to destroy this gap. Furthermore, in this regime, the disorder is suppressed by the interactions, and $D_{f,m}$ goes towards $0$ (and $D_{f,m}\alpha$ doesn't diverge). The same is true for $K_\sigma$, which also goes towards $0$. Since the RG equations are derived perturbatively in $g$, and similarly to what is done for the case of the simple sine-Gordon flow \cite{giamarchi_book_1d}, we stop the flow at $g = 1$, beyond which the RG becomes unreliable. 

On the other hand, if we are above the separatrix, the disorder wins and manages to destroy the spin gap. Both $D_{f,m}$ and $K_\sigma$ go to constants. Even if $g$ diverges, it has no effect on the other parameters of the flow. This regime becomes similar to the study of the previous section Sec.~\ref{sec:g=0} where there was no $g$ term and thus no gap.

\subsubsection{Correlation functions}
We compute using the same renormalization procedure, the correlation function $R_\sigma(r)$. This correlation is given by:
\begin{equation} \label{eq:RG_susceptibility}
\begin{split}
    R_\sigma(r) =&  \frac{1}{2}e^{-K_\sigma\ln(r/\alpha)}e^{-\frac{2K_\sigma^2D_{f,m}}{u_\sigma^2}|x|} e^{-\int_0^{\ln(r/\alpha_o)}\frac{g(l) dl}{\pi u_\sigma}}\\
    &\cdot e^{\int_0^{\ln(r/\alpha_o)}dl\frac{K^2_\sigma(l)g^2(l)}{2\pi^2 u_\sigma^2}\ln(r)F(A(l))}\\
    &\cdot e^{\int_0^{\ln(r/\alpha_o)} dl\frac{4g^2(l)D_{f,m}(l)K^3_\sigma(l)}{\pi^3u_\sigma^4}|x|G(A(l))}\\
    A(l)=&\frac{8D_{f,m}(l)K_\sigma^2(l)}{u_\sigma^2}\alpha(l)
\end{split}
\end{equation}
If $g$ dominates, we have to stop the RG flow when $g$ gets of order $1$. Beyond this point, all the correlations related to the spin degree of freedom are frozen to a constant. So our correlation function starts as a power law like function whose exponent is renormalized as we go to larger scales, until the correlation freezes to a constant.

If the disorder dominates, we do not stop the RG flow, since the diverging $g$ has no effect anymore on the correlation function and the other parameters. Neglecting $g$ is fully justified beyond the point for which $D_{f,m}\alpha \sim O(1)$. Note, however that there is an intermediate regime, due to the term linear in $g$ in the renormalization of $R_\sigma(r)$ that is difficult to control perturbatively, since the renormalized $g$ is large, but the exponential decay that sets in when $D_{f,m}\alpha \sim O(1)$ has not yet started. This case is similar to the perturbative treatment of the commensurate-incommensurate phase transition \cite{horowitz_renormalization_incommensurable}, and ultimately does not affect the physics of the problem. In this regime we thus qualitatively have a correlation function which start as power law, gets corrected a bit (until $g$ stops ``resisting" the disorder) and finally decays exponentially, similarly to its behavior in the model without $g$.

\subsubsection{Critical disorder}

The calculation of the correlation function allows us to determine $D_{fc}$, the critical disorder at which superconductivity is killed. To practically evaluate it, we take in this section the definition that when $T_{min} = 10^{-8}$ in units of $\alpha_o/u_\sigma$ we can consider that this is equivalent to having completely killed the superconductivity. Furthermore, in this case, given the sudden changes in the values of $\chi_o(0,0)$ as we transition from one regime to the other with large interactions, the best that we can numerically do is approach $D_{fc}$ from below.

\begin{figure}
\includegraphics[width=\columnwidth]{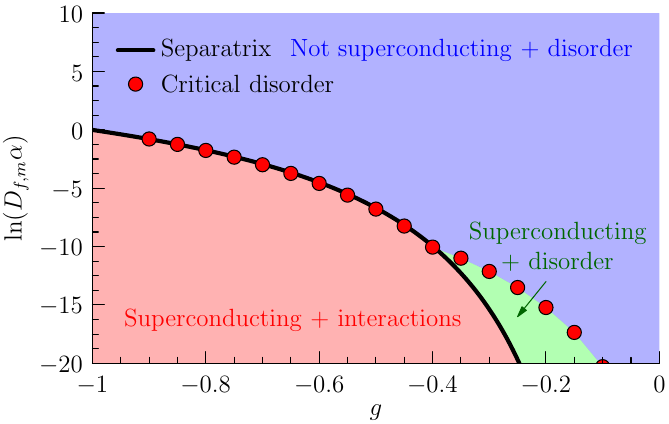}
\caption{\label{fig:separatrix_vs_Dc}
The black line is the separatrix between the 2 RG flow regions, where either the interactions or the disorder dominate. The red points correspond to the critical disorder for a given $U/(g)$, separating the regions with and without superconductivity.}
\end{figure}

In Figure \ref{fig:separatrix_vs_Dc}, we compare the separatrix of our RG flow to the critical disorder for different interactions $U$.
While for small interactions, there exists a region where the system remains superconducting even if disorder dominates the RG flow, for large interactions the critical disorder coincides with the separatrix/ the change of regimes. This can be interpreted as follows : when the spin gap is very small, the competition between disorder and superconductivity is primarily governed by the competition between the random magnetic field, which acts as a random chemical potential for each spin species separately, and the Josephson term that favors a $q=0$ like pairing in a singlet state. When the gap is large, the random magnetic field must first destroy the gap to be effective, but then it very efficiently outcompetes the (small) Josephson coupling term. 

\subsubsection{Critical temperature}

The susceptibility of one chain $\chi_o(q=0,\omega=0)$ is given by: 
\begin{equation}
    \chi_o(0,0)=\int dx \int^{\beta_c}_0 d\tau \; R_\rho(r)R_\sigma(r)
\end{equation}
where $r^2 = x^2+(u_\sigma\tau)^2$ and we neglect the factors of $\alpha$ inside $r$. We use polar coordinates with $u_\sigma\tau=y$. This leads to:
\begin{multline}
    \chi_o(0,0)=\frac{1}{4u_\sigma}\alpha^{K_\sigma+\frac{1}{K_\rho}}\int^{u\beta_c}_{\alpha_o}  drr^{(1-K_\sigma-\frac{1}{K_\rho})}  e^{-\int_0^{\ln(r/\alpha_o)}\frac{g(l)dl}{\pi u_\sigma}}\\
    \cdot e^{\int_0^{\ln(r/\alpha_o)}dl\frac{K^2_\sigma(l)g^2(l)}{2\pi^2 u_\sigma^2}\ln(r)F(A(l)))}\\
    \cdot \int_0^{2\pi}d\theta e^{-\left(\frac{2K_\sigma^2D_{f,m}}{u_\sigma^2}-\int_0^{\ln(r/\alpha_o)}dl\frac{4g^2(l)D_{f,m}(l)K^3_\sigma(l)}{\pi^3u_\sigma^4}G(A(l))\right)|x|}
\end{multline}
After integrating over the angles, we obtain:
\begin{multline}
    \chi_o(0,0)=\frac{1}{4u_\sigma}\alpha^{K_\sigma+\frac{1}{K_\rho}}\int^{u\beta_c}_{\alpha_o}  drr^{1-K_\sigma-\frac{1}{K_\rho}}  \\
    \cdot e^{\int_0^{\ln(r/\alpha_o)}dl(-\frac{g(l)}{\pi u_\sigma}+\frac{K^2_\sigma(l)g^2(l)}{2\pi^2 u_\sigma^2}\ln(r)F(A(l)))}\\
    \cdot 2\pi F\left[\left(\frac{2K_\sigma^2D_{f,m}}{u_\sigma^2}\right.\right.\\
    \left.\left. -\int_0^{\ln(r/\alpha_o)}dl\frac{4g^2(l)D_{f,m}(l)K^3_\sigma(l)}{\pi^3u_\sigma^4}G(A(l))\right)r\right]\\
    \label{eq:chi_0_0_for g_real}
\end{multline}
where the parameters $K_\sigma, D_{f,m}$ outside the integrals over $l$ are the parameters at the beginning of the RG procedure.This expression is similar to the one for the $g=0$ model, with the main differences being the appearance of a linear term in $g$ due to the specific correlation function we are considering and the fact that the parameters $K_\sigma, D_{f,m}$ are here renormalized by the RG flow.

The behaviour of $\chi_o(0,0)$ varies depending on the scale and the regime (interaction or disorder dominated) we are looking. At short scales, in both cases, the integrand of (\ref{eq:chi_0_0_for g_real}) is proportional to the power law $r^{1-\frac{1}{K_\rho}-K_\sigma}$, since $F(r)$ goes to $1$ for small $r$. At large $r$, $\chi_o(0,0)$ behaves differently depending of which parameter dominates. If $g$ dominates, since $R_\sigma(r)$ is now a constant, the integrand of (\ref{eq:chi_0_0_for g_real}) becomes proportional to $r^{1-\frac{1}{K_\rho}}$, and this power law always increases with $r$. Therefore, there is always a finite critical temperature in this regime. On the other hand, if disorder dominates, the integrand of (\ref{eq:chi_0_0_for g_real}) behaves as $r^{-\frac{1}{K_\rho}-K_\sigma}$, which decays rapidly enough to ensure convergence for the interactions considered. This implies that there may not always be a finite $T_c$ in this regime. This abrupt change in behaviour impacts the critical temperature. One should note that this sudden change of behaviour is at least partly due to our treatment in the RG procedure of the two regimes above and below the separatrix as two completely different regimes. This approach allows for a simple estimation of the complicated integral while retaining the main features of the solution. A more complete treatment would smoothen the curve.

\begin{figure}
\includegraphics[width=\columnwidth]{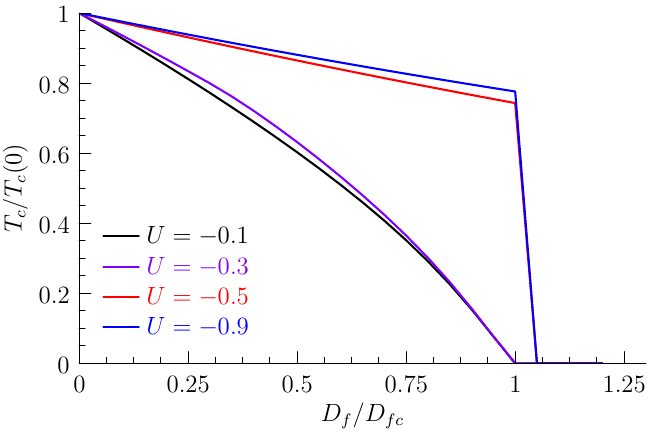}
\caption{\label{fig:Tc_g_neq_0_Dc}
$T_c$ normalized as a function of the strength of magnetic forward disorder normalized over the critical disorder for different values of $U$. 
}
\end{figure}

Finally, we compute the critical temperature by solving (\ref{eq:meandieldeq}). Fig.~\ref{fig:Tc_g_neq_0_Dc} shows two different regimes in the critical temperature. 
To numerically compute those integrals, it is useful to apply different treatments depending on the RG regime. Unlike the case $g=0$, it is better to perform first the integration over the angles $\theta$ and then integrate over $r$. For small enough interactions, we recover a very similar behavior to the $g=0$ case. However, for large interactions, the decrease of $T_c$ is dramatically slowed by the presence of the finite gap, and then $T_c$ drops rapidly to zero, as explained qualitatively above.  

\begin{figure}
\includegraphics[width=\columnwidth]{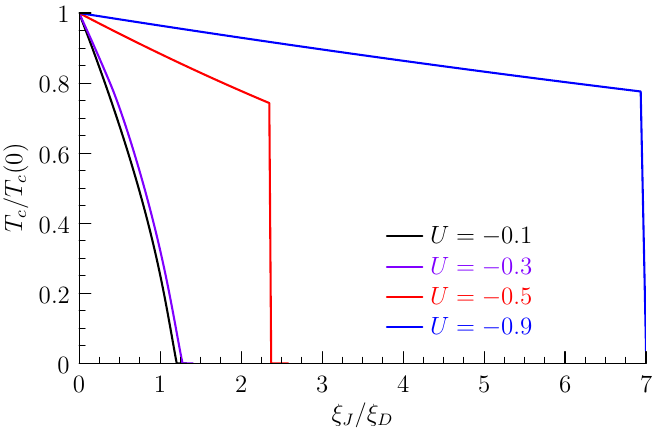}
\caption{\label{fig:Tc_g_neq_0_length}
$T_c$ as a function of the strength of magnetic forward disorder for different values of $U$. The lenghts $\xi_J$ (resp. $\xi_D$) characterize the superconducting order coming from the \emph{transverse coupling $J$} (resp. the exponential decay due to disorder). This distinguishes two regimes depending on whether the spin gap of an \emph{isolated} chain is large or small. }
\end{figure}

Fig.~\ref{fig:Tc_g_neq_0_length} plots the same quantity as a function of the ratio of the two characteristic lengthscales $\xi_J$ (characterizing the superconducting phase in the pure case) and $\xi_D$ (the lengthscale of the exponential decay for the disordered case). As shown in the figure, when the gap is small (small $U$) the superconductivity is suppressed when these two lengths are approximately equal. In contrast, when interactions are large and there is a well formed spin gap, the disorder must first overcome the spin gap, independently of the scale $\xi_J$ which characterize the \emph{transverse} coupling. This leads to the slow and somewhat linear decrease of $T_c$. Once the gap is gone the disorder is at that point large enough to also overcome the contribution coming from the transverse Josephson coupling.

\section{Discussion} \label{sec:Discussion}

\subsection{Comparison with the isotropic BCS solution} \label{sec:BCS_differences}

From the previous section, we see that for the quasi-1D situation, we obtain the equivalent of an Anderson theorem, originally derived close to the $T_c$ in an isotropic situation with a solution of the BCS equations. A weak random chemical potential does not affect $T_c$ or the pairing correlation functions below $T_c$. However, a disorder breaking the time-reversal symmetry, such as a random magnetic disorder, has a more complex effect. Essentially, such a disorder appears in the pair correlation functions and thus will have an effect on the superconductivity as indicated in Fig.~\ref{fig:Tc_g_neq_0_Dc} and Fig.~\ref{fig:Tc_g_neq_0_length}. 

We however consider a situation where $T_c$ and pairing are essentially controlled by the strength of the interchain (Josephson) coupling, with the interactions inside a chain being essentially arbitrary. Thus, it is perfectly possible to have a strong spin gap $\Delta_\sigma$ while having at the same time a small $T_c$. In this case, as shown in Fig.~\ref{fig:Tc_g_neq_0_length}, the magnetic disorder must first destroy the spin gap before it can influence $T_c$ or the pair correlations. For an isotropic case, such a situation would also occur on a lattice, with e.g. an attractive Hubbard model at large $|U|$. In such a situation, the spin gap scale is essentially $|U|$, while the kinetic energy of the pairs becomes $4 t^2/|U|$, leading to a small condensation temperature and thus to a small $T_c$. It would be interesting to study the effect of magnetic disorder in such a system to see if similar effects than the ones observed here would be found.

\subsection{Interchain hopping and Josephson coupling} \label{sec:interjos}

In the model considered in the previous sections, we assumed that the chains were coupled by a Josephson coupling allowing for the hopping of singlet pairs across the chains. As discussed in Sec.~\ref{sec:bosonized_model}, most microscopic realizations actually contain single particle hopping between the chains. In the presence of a spin gap, due to the attractive interaction, the single particle hopping is suppressed and replaced by the Josephson coupling we have considered in this paper. 

For the non-magnetic disorder, keeping only the Josephson coupling poses no problem since the spin gap is preserved by the disorder. The results derived in the previous section are thus directly applicable to systems with single particle hopping, and we do not expect any important difference between the two models. 

For the magnetic disorder, on the contrary, the spin gap is first destroyed by the disorder and it is thus a challenging and important question to know how the results we obtained would apply when the system has single particle hopping to start with. A detailed solution, particularly by renormalization techniques that have been used to tackle the competition between the single particle hopping and the particle-particle or particle-hole hopping \cite{giamarchi_book_1d}, is clearly beyond the scope of the present paper and will be left for a future publication.

However, one can expect the general results derived here with the Josephson coupling to be largely valid. Indeed, the main additional effect for magnetic disorder will be the destruction of the Josephson coupling, leading to singlet pair hopping. This should naively make the destruction of the singlet superconductivity even more efficient than in a model where the Josephson coupling is kept constant. We can thus expect naively an even stronger effect of the magnetic disorder, making the contrast between the magnetic and non-magnetic disorder even more marked. Another interesting possibility when looking at a model containing single-particle tunneling is that there might also appear some intermediate region in disorder, where the reduction of the spin-gap due to the disorder could lead to an increase in the coupling $J$. This increase would ``oppose" the decrease of the pair-susceptibility and therefore might increase $T_c$.

An interesting possibility for the case of single particle hopping is the potential to stabilize other phases in presence of the magnetic disorder. One order parameter that would be robust to the magnetic disorder is 
\begin{equation}
 O_{TS,xy}(r) \sim e^{i\sqrt2 \theta_\rho(r)} \cos\left(\sqrt2 \theta_\sigma(r)\right) 
 \end{equation}
or the equivalent one with a sine. This order parameter corresponds to the $x$ or $y$ component of a triplet order parameter. A corresponding pair-hopping term is also generated by the single particle hopping but is subdominant for an attractive interaction. Indeed, since the field $\phi_\sigma$ orders, the correlations of the field $\theta_\sigma$ decrease exponentially fast. However, in presence of the magnetic disorder, the $\cos(\sqrt8 \phi_\sigma)$ term in the single chain Hamiltonian (\ref{eq:singlechain}) is essentially killed, and the $\theta_\sigma$ are the only correlations without exponential decay. Since $K_\rho > 1$ due to the attractive interaction, the $\theta_\rho$ correlations are favored in the charge sector. 

This would lead to the interesting possibility of replacing the singlet superconducting phase with a triplet one when the magnetic disorder becomes large enough. Of course, this phase will have a lower $T_c$ but should survive even at relatively large magnetic disorder. This could be a practical possibility to stabilize a triplet superconducting phase even in the presence of contact attractive interaction, e.g. in a cold atom realization.

\subsection{Possible implementations} \label{sec:implementations}

To test for the the effects investigated here, cold atomic systems provide a natural potential realization. Several key ingredients needed could be realized in such systems. The coupled 1D structures of fermions with attrative interactions can be readily realized, either in systems made of several tubes \cite{hart_antiferro_Hubbard_ultracold_atoms} or in systems with quantum microscopes \cite{boll_quantum_microscope_spin_resolved,bakr_gillen_09}. 

One of the required key ingredients is a disorder that would be mainly forward. This also can be realized either with the natural limitation provided by a speckle disorder \cite{lugan_correlated_potentials} or in systems such as quantum microscopes by generating the disorder via DMD and tuning the Fourier transform of the disorder so that Fourier components close to $2 k_F$ are absent. 

Measuring $T_C$ itself might not be experimentally easy to realize, but a simpler measurement could be provided by the decay of the pair correlation functions along the tubes, which are a direct measure of the existence of superconductivity in the system. In that respect, for quantum microscopes, since the easily measured quantity is the density, it could be useful to make use of the relation that exists between the attractive and repulsive Hubbard models \cite{ho_attractive_hubbard}. Such a transformation maps the attractive model into the repulsive one and the random chemical potential into a random magnetic field and vice versa. The observables are directly related by a particle-hole transformation on spin down only \cite{ho_attractive_hubbard}. In particular, a singlet order parameter would map onto an antiferromagnetic spin order along the $x$ or $y$ direction. 

In the language of the repulsive Hubbard model, one would thus conclude from the results of the previous sections that a \emph{random magnetic field along $z$} would essentially not affect the correlation function of the antiferromagnetic fluctuations along $x$ or $y$ of a half filled system (one particle per site), leading to a $T_C$ for antiferromagnetic order in the plane that is essentially unchanged. This corresponds to the Anderson theorem for the non-magnetic disorder. On the other hand, putting a random chemical potential along the tubes would drastically destroy such correlation, corresponding to the destructive effects of the \emph{magnetic} disorder on single superconductivity that we have found in the present study. The competition discussed earlier between the magnetic disorder and the spin gap in the attractive side becomes the competition between the Mott gap and the random chemical potential on the repulsive side. It is necessary for the random chemical potential to be stronger than the Mott gap to locally dope the system. Once this is reached, however, the \emph{spin-spin} correlation in the $x-y$ plane gets very rapidly destroyed.

In condensed matter systems, one would have to use highly anisotropic systems. Organic superconductors are a good candidate \cite{jerome_comptes_rendus}, and there may also be a possibility of investigation in some 2D systems where high anisotropy has been reported, such as CrSBr \cite{fan_CrSBr}. The easy part here is identifying superconductivity since simple resistance measurement (and Meissner effect) are routine experiments. However, the hard part in this kind of experiments would be controlling the disorder to keep the backscattering much smaller than the forward scattering along the chains. One possibility would be, like for two dimensional semiconducting systems, to place the impurities far from the conducting chains. This, however, would have the drawback to also lead to quite correlated potentials from one chain to the next. 

\section{Conclusion} \label{sec:conclusion}
In this work, we extended the Anderson theorem for superconductivity, which states that non-magnetic impurities do not impact a BCS superconductor, meaning they do not change its critical temperature, while magnetic impurities have a drastic effect. We considered a quasi-1d system with forward scattering disorder, coupled by a Josephson coupling favoring singlet superconductivity. We showed that for such a system, the non-magnetic forward disorder leaves $T_C$ and pair correlations essentially unchanged. On the other hand, magnetic disorder has a significant impact on the system. Once such disorder overcomes the spin-gap, it starts destroying the pair correlations and hence the superconductivity very efficiently. Interestingly, the correlation function that seems to survive the magnetic disorder and still decays slowly is the $xy$ part of the triplet superconducting correlation (with a random magnetic field along $z$). 

We also discussed various possible tests of these predictions in condensed matter and especially in cold atomic gases. Quantum microscopes provide an ideal system to test for the predictions of this paper, using an implementation for \emph{a repulsive} Hubbard model with one particle per site. In that case, magnetic disorder would leave the $xy$ antiferromagnetic spin-spin correlations essentially unchanged, while a \emph{non-magnetic} disorder would rapidly destroy such correlations once it is able to suppress the Mott gap. 

Several extentions of our work would be interesting. In particular, since chains are coupled by single-particle tunnelling and not just by the pair tunnelling, other pair couplings can be generated. This raises to the question of which instability could be dominant once the singlet superconductivity has been destroyed. The most likely candidate is a triplet superconducting pairing. Whether such pairing, which dominates for a single chain, could be effectively stabilizes in the 2D or 3D case is an intereting question and challenge. Indeed if this is the case, it would provide a route to realize triplet superconducting phases with purely contact interactions. These questions will be examined in future 
studies. 

\acknowledgments
We are thankful to Christophe Berthod for his help in the numerical solution of our equations and for very useful discussions. This research was supported by the Swiss National Science Foundation under grants 200020-188687 and 200020-219400.

\appendix
\section{Computing the RG equations} \label{app:RG_equations}

In this appendix, we describe in more details how to compute $H_\sigma(r_a)$ and derive the RG equations (\ref{eq:RG_K}, \ref{eq:RG_g}, \ref{eq:RG_disorder}, \ref{eq:RG_susceptibility}).

We start from : 
\begin{multline}    H_\sigma(r_a)=\frac{1}{Z_{\phi_{\sigma}}}e^{\frac{2K_\sigma^2 D_{f,m}|x_a|}{u_\sigma^2}}e^{\frac{K_\sigma}{2}\ln \left(\frac{x_a^2+(u_\sigma|\tau_a|+\alpha)^2}{\alpha^2}\right)}\\
    \int D\phi_\sigma \sum_{\varepsilon_1,\varepsilon_2 = \pm 1}\frac{1}{4}e^{i\varepsilon_1\sqrt{2}\phi_\sigma(r_a)}e^{i\varepsilon_2\sqrt{2}\phi_\sigma(0)}e^{-S_{\phi_\sigma}}
\end{multline}
where $S_{\phi_\sigma}$ is the full 1d action of the spin sector after integrating out the $\theta$ degrees of freedom, and $Z_\phi{_\sigma}$ is the partition function associated to it. We first absorb the disorder terms of the Hamiltonian in the definition of the field $\phi_\sigma$ and replace them by $\tilde{\phi}_\sigma$ (see also the main text). We then expand the action to the 2nd order in $g$. For simplicity of notation, we will drop the $\sigma$ of $\phi_\sigma$ and the $m$ of $D_{f,m}$ in the following equations.

Integrating over the configurations leads to the expansion in powers of $g$:
\begin{widetext}
\begin{align*}
H_\sigma(r_a)& = e^{\frac{2K_\sigma^2 D_{f}|x_a|}{u_\sigma^2}}e^{\frac{K_\sigma}{2}\ln \left(\frac{x_a^2+(u_\sigma|\tau_a|+\alpha)^2}{\alpha^2}\right)} \\
\cdot &\left[\sum_{\varepsilon=\pm 1}\frac{1}{4}e^{-i\varepsilon\frac{2K_\sigma}{u_\sigma}\int_0^{x_a}\gamma(x')dx'} \langle e^{i\varepsilon\sqrt{2}(\tilde{\phi}(r_a)-\tilde{\phi(0)})}\rangle_{H_{o\tilde{\phi}}} \right.\\
& -\frac{g}{8\pi^2\alpha^2}\int dx d\tau \sum_\varepsilon e^{-i\varepsilon\frac{2K_\sigma}{u_\sigma}\left(\int_x^{x_a}+\int_x^0\right)\gamma(x')dx'}\langle e^{i\varepsilon\sqrt{2}(\tilde{\phi}(r_a)+\tilde{\phi(0)})-2\tilde{\phi}(x)}\rangle_{H_{o\tilde{\phi}}}\\
&+\frac{1}{16}\frac{g^2}{8\pi^4\alpha^4}\int dx_1 d\tau_1 dx_2 d\tau_2\sum_{\varepsilon_1,\varepsilon_2}e^{-i\frac{2K_\sigma}{u_\sigma}\left(\varepsilon_1\int_0^{x_a}+2\varepsilon_2 \int_{x_2}^{x_1}\right)\gamma(x')dx'} \langle e^{i\varepsilon_1\sqrt{2}(\tilde{\phi}(r_a)-\tilde{\phi(0)})}e^{i\varepsilon_2\sqrt{8}(\tilde{\phi}(x_1)-\tilde{\phi(x_2)})}\rangle_{H_{o\tilde{\phi}}} \\
&\left. -\frac{1}{16}\frac{g^2}{8\pi^4\alpha^4}\int dx_1 d\tau_1 dx_2 d\tau_2\sum_{\varepsilon_1,\varepsilon_2}e^{-i\frac{2K_\sigma}{u_\sigma}\left(\varepsilon_1\int_0^{x_a}+2\varepsilon_2 \int_{x_2}^{x_1}\right)\gamma(x')dx'} \langle e^{i\varepsilon_1\sqrt{2}(\tilde{\phi}(r_a)-\tilde{\phi(0)})}\rangle_{H_{o\tilde{\phi}}} \langle e^{i\varepsilon_2\sqrt{8}(\tilde{\phi}(x_1)-\tilde{\phi(x_2)})}\rangle_{H_{o\tilde{\phi}}}\right]
\end{align*}
\end{widetext}
The average on $H_{o\tilde{\phi}}$ is an average on the quadratic part of the Hamiltonian expressed in terms of $\tilde{\phi}$.

\subsection{Disorder averages}
To perform the disorder averages in the above expression, we rewrite the integrals with the help of the Heaviside function. Completing the square and simplifying the result leads for the first disorder average to:
\begin{align*}
    \frac{1}{Z_D}\int d\gamma e^{-\frac{1}{2D_{f}} \int dx' \gamma(x')^2-i\frac{2K_\sigma\varepsilon}{u_\sigma}\int_0^{x_a}\gamma(x')dx'} = e^{-\frac{2D_fK_\sigma^2}{u_\sigma^2}|x_a|}
\end{align*}
where $Z_D = \int d\gamma e^{-\frac{1}{2D_{f,m}} \int dx' \gamma(x')^2}$.

The second disorder average (linear term in $g$) similarly leads to:
\begin{multline}
    e^{\frac{2K_\sigma^2D_f}{u_\sigma^2}\left[|x_a-x|+|x|\right]} \cdot \\ 
    e^{-2\left([\min(x_a,0-x)]\theta(-x)\theta(x_a-x) +[x-\max(0,x_a)]\theta(x)\theta(x-x_a)\right)}
\end{multline}

For the third term (second order in $g$), we obtain:
\begin{align}
\label{eq:app_average_disorder_second_order_term}
    \nonumber&e^{\frac{-2K_\sigma^2 D_f}{u_\sigma^2}|x_a|-\frac{-8K_\sigma^2 D_f}{u_\sigma^2}|x_1-x_2|}\\
    \nonumber&\cdot e^{-\varepsilon_1\varepsilon_2\frac{8D_fK_\sigma^2}{u_\sigma^2}\left(\min(x_1,x_a)-\max(0,x_2)\right)\theta(x_a)\theta(x_1)\theta(\min(x_1,x_a)-x_2)}\\
    \nonumber&\cdot e^{-\varepsilon_1\varepsilon_2\frac{8D_fK_\sigma^2}{u_\sigma^2}\left(\min(0,x_2)-\max(x_1,x_a)\right)\theta(-x_a)\theta(-x_1)\theta(x_2-\max(x_1,x_a))}\\
    \nonumber&\cdot e^{\varepsilon_1\varepsilon_2\frac{8D_fK_\sigma^2}{u_\sigma^2}\left(\min(x_2,x_a)-\max(0,x_1)\right)\theta(x_a)\theta(x_2)\theta(\min(x_2,x_a)-x_1)}\\
     &\cdot e^{\varepsilon_1\varepsilon_2\frac{8D_fK_\sigma^2}{u_\sigma^2}\left(\min(0,x_1)-\max(x_2,x_a)\right)\theta(-x_a)\theta(-x_2)\theta(x_1-\max(x_2,x_a))}
\end{align}

\subsection{Further algebra and some tricks}
The remaining configuration averages can be easily computed since $H_{o\tilde{\phi}}$ is quadratic. In this case, we use the fact that $\langle e^A\rangle = e^{\frac{1}{2}\langle A^2\rangle}$ and the relation \cite{giamarchi_book_1d} 
\begin{align}
    \nonumber\langle (\phi(r_1)-\phi(r_2))^2\rangle &= \frac{K_\sigma}{2}\ln\left[\frac{x^2+(u_\sigma\tau+\alpha)^2}{\alpha^2}\right]\\
    &=K_\sigma F_1(r_1-r_2)
\end{align}
We can then combine the connected and disconnected terms of the second order expansion since they are similar in all of their terms expect for the cross terms between $(0,x_a)$ and $(x_1,x_2)$ which appear only in the connected term. This leads to a term of the form:
\begin{align}
\label{eq:app_equation_leading_interaction_RG}
    \dots \left[e^{2\varepsilon_1\varepsilon_2K_\sigma (F_1(r_a-r_1)-F_1(r_a-r_2)-F_1(0-r_1)+F_1(0-r_2))}-1\right]
\end{align}

From (\ref{eq:app_average_disorder_second_order_term}) we obtain: 
\begin{widetext}
\begin{equation} \label{eq:app_separation_disorder_interaction}
\begin{split}
    &e^{\frac{-2K_\sigma^2 D_f}{u_\sigma^2}|x_a|-\frac{-8K_\sigma^2 D_f}{u_\sigma^2}|x_1-x_2|} \cdot\\
    &\left[\left(e^{-\varepsilon_1\varepsilon_2\frac{8D_fK_\sigma^2}{u_\sigma^2}\left(\min(x_1,x_a)-\max(0,x_2)\right)\theta(x_a)\theta(x_1)\theta(\min(x_1,x_a)-x_2)}\right.\right.
     e^{-\varepsilon_1\varepsilon_2\frac{8D_fK_\sigma^2}{u_\sigma^2}\left(\min(0,x_2)-\max(x_1,x_a)\right)\theta(-x_a)\theta(-x_1)\theta(x_2-\max(x_1,x_a))} \cdot\\
    &\cdot e^{\varepsilon_1\varepsilon_2\frac{8D_fK_\sigma^2}{u_\sigma^2}\left(\min(x_2,x_a)-\max(0,x_1)\right)\theta(x_a)\theta(x_2)\theta(\min(x_2,x_a)-x_1)}
      \left.\left.e^{\varepsilon_1\varepsilon_2\frac{8D_fK_\sigma^2}{u_\sigma^2}\left(\min(0,x_1)-\max(x_2,x_a)\right)\theta(-x_a)\theta(-x_2)\theta(x_1-\max(x_2,x_a))} -1\right) + 1\right]
\end{split}
\end{equation}
\end{widetext}
which allows us to obtain the renormalization equations for the interaction terms and for the disorder separately.

\subsection{The interaction renormalization equations}
We apply the procedure described in \cite{giamarchi_book_1d} to the second term of (\ref{eq:app_separation_disorder_interaction}) multiplied by (\ref{eq:app_equation_leading_interaction_RG}). In this part of our equations, no disorder ($D_f$) appears. Rewriting $r_1$ and $r_2$ as center of mass ($R = (X=(x_1+x_2)/2, Y=u_\sigma (\tau_1+\tau_2)/2)$) and relative coordinates $r$, we recognize gradients of $F_1$ : ($\nabla_R F_1(r_a-R)$). In the expansion in small $r$ up to second order, the $\nabla^2_X-\nabla^2_Y$ term renormalizes the velocity $u_\sigma$. We neglect this contribution since the change of velocity does not affect the physics of the problem in an essential way. On the contrary, we retain $\nabla^2_X+\nabla^2_Y$, which can be simplified using $(\nabla^2_X+\nabla^2_Y)\log(R)=2\pi\delta(R)$. 

Using polar coordinates the full term becomes: 
\begin{align}
    \frac{K_\sigma^2g^2}{4\pi^2\alpha^4u_\sigma^2}e^{-\frac{2D_fK_\sigma^2}{u_\sigma^2}|x_a|}F_1(r_a) \int_{r>\alpha}dre^{-4K_\sigma F_1(r)}r^3F\left(\frac{8D_fK_\sigma^2}{u_\sigma^2}r\right)
\end{align}
where $F$ has been defined in (\ref{eq:definition_F})

Looking at the contribution of $\int_\alpha^{\alpha+d\alpha}dx$ where we define $\alpha = \alpha_o e^l$, we obtain the RG equations for $K_\sigma$ (\ref{eq:RG_K}) and $g$ (\ref{eq:RG_g}).

\subsection{Disorder renormalization equations}
We now look at the first term of (\ref{eq:app_separation_disorder_interaction}) multiplied by (\ref{eq:app_equation_leading_interaction_RG}). We discuss the case $x_a>0$ since the other case can be treated similarly. Since for for large disorders, the 2 exponentials $e^{\frac{-2K_\sigma^2 D_f}{u_\sigma^2}|x_a|-\frac{-8K_\sigma^2 D_f}{u_\sigma^2}|x_1-x_2|}$ which we don't expand would suppress completely this term, we can expand the exponentials containing the ``crossed" disorder terms and the exponential contaning the $F_1$ terms at first order.

We now also have to separate the integrals on $x_1$ and $x_2$ to handle all cases arising from the Heaviside functions. Those split in 4 categories (our chains have size $2L$): $\int^{x_a}_0 dx_1\int^{x_1}_0 dx_2$, $\int^{x_a}_0 dx_1\int^{0}_{-L} dx_2$, $\int^{L}_{x_a} dx_1\int^{x_a}_{-L} dx_2$, $\int^{L}_{x_a} dx_1\int^{0}_{-L} dx_2$.

Since the term is suppressed exponentially in $|x_1-x_2|$, the fourth category of integrals is negligible since the minimal interval between $x_1$ and $x_2$ is $L$, which is half the size of the system. The most important contribution comes from from the first term, which covers most of the $|x_1-x_2|$ ``small" region, and gives us the renormalization that we use. Finally, the second and third term are ``marginally relevant" in the sense that while they have also possibilities of having small $|x_1-x_2|$, they have only one point where $x_1$ and $x_2$ coincide. This last terms would give rise to less relevant terms in the renormalization equations. 

Another way of looking at it is that we are looking at the effect of having disorder on two points, $x_1$ and $x_2$, on the correlations related to two fixed point, $x_a$ and $0$. Since the effect of $x_1$ and $x_2$ is suppressed if they are far from each other, given the splitting of our integral, the importance of each contribution is related to the amount of possibilities to have $x_1$ and $x_2$ close together.

We then go to the center of mass and relative coordinates and perform first the integral on the center of mass coordinates first, starting with the integral on the center of mass "time". We are now looking at:
\begin{multline}
    16\frac{8D_fK_\sigma^3}{u_\sigma^4} \int dy \int_0^{x_a} dx e^{-\frac{8D_fK_\sigma^2}{u_\sigma^2}|x|}e^{-4K_\sigma F_1(r)}x^2\\
     \cdot \int_{x_a-x/2}^{x/2} dX \int dY \nabla_X\left[F_1(-R)-F_1(r_a-R)\right]
\end{multline}
which then leads to:
\begin{equation}
    16\pi\frac{8D_fK_\sigma^3}{u_\sigma^4}\int dy \int_0^{x_a} dx e^{-\frac{8D_fK_\sigma^2}{u_\sigma^2}|x|}e^{-4K_\sigma F_1(r)}x^2(x_a-x)
\end{equation}

After converting to polar coordinates and performing the integration, we end up with:
\begin{multline}
   16\pi\frac{8D_fK_\sigma^3}{u_\sigma^4} \left[\int_0^{x_a}dr r^3 e^{-4K_\sigma F_1(r)}x_a G\left(\frac{8D_f K_\sigma^2}{u_\sigma^2}r\right)\right.\\
    \left.-\int_0^{x_a} dr r^4 e^{-4K_\sigma F_1(r}M\left(\frac{8D_f K_\sigma^2}{u_\sigma^2}r)\right)\right]   
\end{multline}
with
\begin{equation}
 M(x) = \frac{4}{3}-\pi(I_3(x)-L_1(x))-\pi\frac{I_2(x)-L_2(x)}{x}
\end{equation}

In this last expression, the second term leads to a new RG equation which is less relevant. The first term instead, when re-exponentiated leads to the RG equation for the disorder 
(\ref{eq:RG_disorder}). 

\subsection{Correlation function renormalization}
The last element that we need comes from the linear term in $g$. By splitting the integral on $dx$ to handle the Heaviside functions, we find that the disorder term can be rewritten uniformly as: $e^{\frac{2K_\sigma^2D_f}{u_\sigma^2}|x_a|}$. Finally, combining all of this terms together, we derive the RG equation of the correlation function (\ref{eq:RG_susceptibility}).

\section{Details on the interpretation of the RG equations}
\label{sec:app_interpretation_RG_equation}
We can play around a bit with our RG equations (\ref{eq:RG_K}), (\ref{eq:RG_g}), (\ref{eq:RG_disorder}) by noticing that in all of them the disorder strength $D_{f,m}$ is accompanied by $\alpha$.  This suggests making a change of variable $\tilde{D}=D_{f,m}\alpha$ (the quantity which we compare to $g$ to decide who wins the RG).

This leads to the following RG equations :
\begin{equation}
\begin{split}
    \frac{dK_\sigma}{dl}&=-\frac{g^2K_\sigma^2}
    {2\pi^2u_\sigma^2}F\left(\frac{8K_\sigma^2 \tilde{D}}{u_\sigma^2}\right)\\
    \frac{dg}{dl}&=(2-2K_\sigma)g\\
    \frac{d\tilde{D}}{dl}&=\left(1-\frac{g^2K_\sigma}{\pi^3u_\sigma^2}G\left(\frac{8K_\sigma^2 \tilde{D}}{u_\sigma^2}\right)\right)\tilde{D}
\end{split}
\end{equation}
In this form the competition between $g$ and $\tilde{D}$ is evident. Both would be diverging exponentials if the other is set to $0$ and both are contained (or even suppressed for the equation of $\tilde{D}$) by the other. The main question which remains at this stage is which one of the two will diverge first. 

Since $F$ and $G$ are complicated functions which have simple power law behaviours at large argument, we can also expand them at large argument. Particular attention should be given that this means that $\tilde{D}$ is large, and therefore $D\neq 0$. So this expansion describes well the case where the disorder wins while one has to be more careful if $g$ wins the RG. We then get the following expressions for the RG equations :
 \begin{equation}
 \begin{split}
    \frac{dK_\sigma}{dl} &= -\frac{g^2}{8\pi^3 D_{f,m}}\frac{1}{\alpha}\\
    \frac{dg}{dl} &= (2-2K_\sigma)g\\
    \frac{dD_{f,m}}{dl} &= -\frac{2g^2u_\sigma^4 }{8^3\pi^3D_{f,m}^2K_\sigma^5}\frac{1}{\alpha^3}
\end{split}
\end{equation}
where $\alpha=\alpha_o e^l$. Here we can see clearly the competition between disorder and $g$ if we do the same variable change as just above. However, another change of variable illustrates another aspect of our RG. If we redefine our parameter $g\to \tilde{g}=\frac{g}{\sqrt{\alpha}}$, we get for the RG equations :
\begin{equation}
\begin{split}
    \frac{dK_\sigma}{dl} &= -\frac{\tilde{g}^2}{8\pi^3 D_{f,m}}\\
    \frac{d\tilde{g}}{dl} &= (3/2-2K_\sigma)\tilde{g}\\
    \frac{dD_{f,m}}{dl} &= -\frac{2\tilde{g}^2u_\sigma^4 }{8^3\pi^3D_{f,m}^2K_\sigma^5}\frac{1}{\alpha^2}
\end{split}
\end{equation}
Since this expressions are valid at large $D_{f,m}\alpha$, the disorder equations shows us that the parameter $D_{f,m}$ is in practice frozen. But the main interest here is the second equation, the one for $\tilde{g}$. We see that $\tilde{g}$ is relevant for $K_\sigma>0.75$, which is reminiscent of the case of a backward disorder that would exist only in the spin sector \cite{giamarchi_book_1d}. 

\section{Different velocities}
\label{sec:app_diff_velocities}

\subsection{Model $g=0$}
For the following, we treat here the charge velocity and spin velocity as different.
The susceptibility of one chain $\chi_o(q=0,\omega=0)$ is then given by :
\begin{equation}
    \chi_o(0,0)=\int dx \int^{\beta_c}_0 d\tau R_\rho(r_\rho)R_\sigma(r_\sigma)
\end{equation}
where $r_\rho^2 = x^2+(u_\rho\tau)^2$ and  $r_\sigma^2 = x^2+(u_\sigma\tau)^2$, we neglect the factors of $\alpha$ inside $r_{\rho/\sigma}$. We make the change of variables $u_\sigma\tau=y$, which when passing in polar coordinates leads us to :
\begin{multline}
    \chi_o(0,0)=\frac{1}{4u_\sigma}\alpha^{1+\frac{1}{K_\rho}}\int^{u_\sigma \beta_c}_{\alpha_o}  drr^{-\frac{1}{K_\rho}}\\
    \cdot \int_0^{2\pi}d\theta \frac{e^{-\frac{2K_\sigma^2D_{f,m}}{u_\sigma^2}|x|}}{\left(\cos(\theta)^2+\frac{u_\rho}{u_\sigma}^2\sin(\theta)^2\right)^\frac{1}{K_\rho}}
\end{multline}

\begin{figure}
\includegraphics[width=\columnwidth]{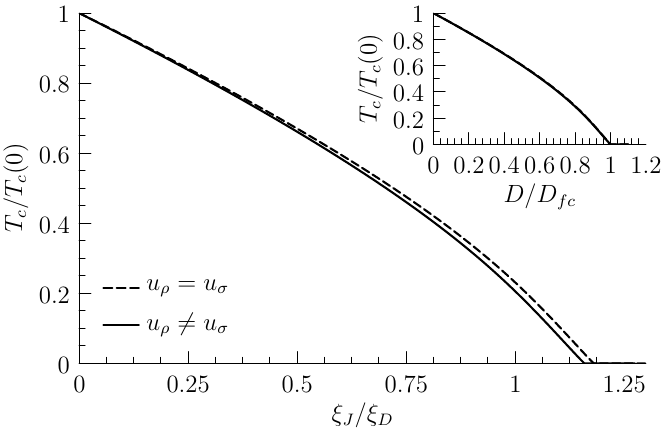}
\caption{\label{fig:app:compare_u_treatment_g=0}
Plots of the behavior of the critical temperature by respect of disorder comparing both treatments of $u_\rho/u_\sigma$ as described in the main text and in this appendix.
}
\end{figure}

As we can see in Fig.~\ref{fig:app:compare_u_treatment_g=0}, the difference between the cases where we consider $u_\sigma = u_\rho$ or where we take their real values coming from their definition (\ref{eq:Hubbard_parameters}) is minimal. So the qualitative behavior can be quite well described analytically by the expression (\ref{eq:selfgzero}) without taking into account the difference of the speeds, while if one wants to be slightly more quantitative, one can refer to the expression tracking the difference between $u_\rho$ and $u_\sigma$.

\subsection{Model $g \neq 0$}

In the same spirit as above, the susceptibility is now given as a function of two different $r$ : $r_\rho$ and $r_\sigma$.
\begin{equation}
    \chi_o(0,0)=\int dx \int^{\beta_c}_0 d\tau R_\rho(r_\rho)R_\sigma(r_\sigma)
\end{equation}
Which then lead by the same change of variables $u_\sigma\tau=y$ to :
\begin{multline}
    \chi_o(0,0)=\frac{1}{4u_\sigma}\alpha^{K_\sigma+\frac{1}{K_\rho}}\int^{u_\sigma\beta_c}_{\alpha_o}  drr^{1-K_\sigma-\frac{1}{K_\rho}}  e^{-\int_0^{\ln(r/\alpha_o)}\frac{gdl}{\pi u_\sigma}}\\
    \cdot e^{\int_0^{\ln(\frac{r}{\alpha_o})}dl \frac{K_\sigma^2g^2}{2\pi^2 u_\sigma^2}\ln(r)F(A(l)))}\\
    \cdot \int_0^{2\pi}d\theta \frac{e^{-\left(\frac{2K_\sigma^2D_{f,m}}{u_\sigma^2}-\int_0^{\ln(r/\alpha_o)}dl\frac{4g^2D_{f,m}K_\sigma^3}{\pi^3u_\sigma^4}G(A(l))\right)|x|}}{\left(\cos(\theta)^2+\frac{u_\rho}{u_\sigma}^2\sin(\theta)^2\right)^\frac{1}{K_\rho}}
\end{multline}

\begin{figure}
\includegraphics[width=\columnwidth]{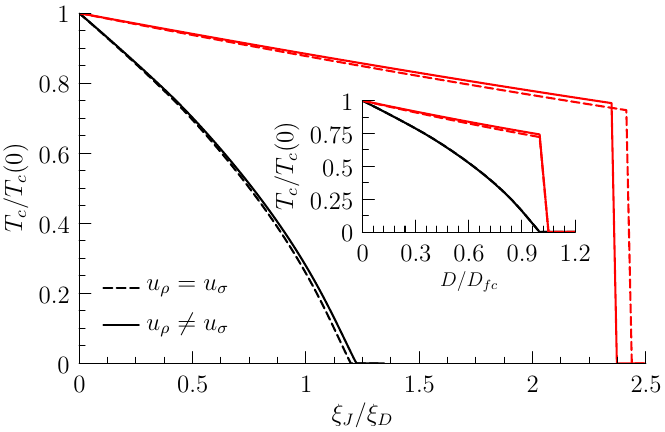}
\caption{\label{fig:app:compare_u_treatment_g_real}
Plots of the behaviour of the critical temperature by respect of disorder comparing both treatments of $u_\rho/u_\sigma$ as described in the main text and in this appendix in both regimes of the RG. 
}
\end{figure}
and we can again see in Figure \ref{fig:app:compare_u_treatment_g_real}, that the qualitative behaviour is the same in both treatments of the speeds for both regimes of the RG.


\begin{thebibliography}{42}%
\makeatletter
\providecommand \@ifxundefined [1]{%
 \@ifx{#1\undefined}
}%
\providecommand \@ifnum [1]{%
 \ifnum #1\expandafter \@firstoftwo
 \else \expandafter \@secondoftwo
 \fi
}%
\providecommand \@ifx [1]{%
 \ifx #1\expandafter \@firstoftwo
 \else \expandafter \@secondoftwo
 \fi
}%
\providecommand \natexlab [1]{#1}%
\providecommand \enquote  [1]{``#1''}%
\providecommand \bibnamefont  [1]{#1}%
\providecommand \bibfnamefont [1]{#1}%
\providecommand \citenamefont [1]{#1}%
\providecommand \href@noop [0]{\@secondoftwo}%
\providecommand \href [0]{\begingroup \@sanitize@url \@href}%
\providecommand \@href[1]{\@@startlink{#1}\@@href}%
\providecommand \@@href[1]{\endgroup#1\@@endlink}%
\providecommand \@sanitize@url [0]{\catcode `\\12\catcode `\$12\catcode
  `\&12\catcode `\#12\catcode `\^12\catcode `\_12\catcode `\%12\relax}%
\providecommand \@@startlink[1]{}%
\providecommand \@@endlink[0]{}%
\providecommand \url  [0]{\begingroup\@sanitize@url \@url }%
\providecommand \@url [1]{\endgroup\@href {#1}{\urlprefix }}%
\providecommand \urlprefix  [0]{URL }%
\providecommand \Eprint [0]{\href }%
\providecommand \doibase [0]{https://doi.org/}%
\providecommand \selectlanguage [0]{\@gobble}%
\providecommand \bibinfo  [0]{\@secondoftwo}%
\providecommand \bibfield  [0]{\@secondoftwo}%
\providecommand \translation [1]{[#1]}%
\providecommand \BibitemOpen [0]{}%
\providecommand \bibitemStop [0]{}%
\providecommand \bibitemNoStop [0]{.\EOS\space}%
\providecommand \EOS [0]{\spacefactor3000\relax}%
\providecommand \BibitemShut  [1]{\csname bibitem#1\endcsname}%
\let\auto@bib@innerbib\@empty
\bibitem [{\citenamefont {Abrikosov}\ and\ \citenamefont
  {Gorkov}(1961)}]{Abrikosov1961}%
  \BibitemOpen
  \bibfield  {author} {\bibinfo {author} {\bibfnamefont {A.}~\bibnamefont
  {Abrikosov}}\ and\ \bibinfo {author} {\bibfnamefont {L.}~\bibnamefont
  {Gorkov}},\ }\bibfield  {title} {\bibinfo {title} {Contribution to the theory
  of superconducting alloys with paramagnetic impurities},\ }\href@noop {}
  {\bibfield  {journal} {\bibinfo  {journal} {Journal of Experimental and
  Theoretical Physics}\ }\textbf {\bibinfo {volume} {12}},\ \bibinfo {pages}
  {1243} (\bibinfo {year} {1961})}\BibitemShut {NoStop}%
\bibitem [{\citenamefont {Anderson}(1959)}]{anderson_dirty_superconductor}%
  \BibitemOpen
  \bibfield  {author} {\bibinfo {author} {\bibfnamefont {P.~W.}\ \bibnamefont
  {Anderson}},\ }\bibfield  {title} {\bibinfo {title} {Theory of dirty
  superconductors},\ }\href
  {https://doi.org/https://doi.org/10.1016/0022-3697(59)90036-8} {\bibfield
  {journal} {\bibinfo  {journal} {Journal of Physics and Chemistry of Solids}\
  }\textbf {\bibinfo {volume} {11}},\ \bibinfo {pages} {26} (\bibinfo {year}
  {1959})}\BibitemShut {NoStop}%
\bibitem [{\citenamefont {Varma}(2006)}]{varma_review_heavy_fermions}%
  \BibitemOpen
  \bibfield  {author} {\bibinfo {author} {\bibfnamefont {C.}~\bibnamefont
  {Varma}},\ }\bibfield  {title} {\bibinfo {title} {Thirty years of heavy
  fermions: Scientific setting for their discovery and partial understanding},\
  }\href {https://doi.org/https://doi.org/10.1016/j.physb.2006.01.019}
  {\bibfield  {journal} {\bibinfo  {journal} {Physica B: Condensed Matter}\
  }\textbf {\bibinfo {volume} {378-380}},\ \bibinfo {pages} {17} (\bibinfo
  {year} {2006})},\ \bibinfo {note} {proceedings of the International
  Conference on Strongly Correlated Electron Systems}\BibitemShut {NoStop}%
\bibitem [{\citenamefont {Jerome}\ and\ \citenamefont
  {Bourbonnais}(2024)}]{jerome_comptes_rendus}%
  \BibitemOpen
  \bibfield  {author} {\bibinfo {author} {\bibfnamefont {D.}~\bibnamefont
  {Jerome}}\ and\ \bibinfo {author} {\bibfnamefont {C.}~\bibnamefont
  {Bourbonnais}},\ }\bibfield  {title} {\bibinfo {title} {Quasi one-dimensional
  organic conductors: from fröhlich conductivity and peierls insulating state
  to magnetically-mediated superconductivity, a restrospective},\ }\href
  {https://doi.org/10.5802/crphys.164} {\bibfield  {journal} {\bibinfo
  {journal} {Comptes Rendus de Physique}\ }\textbf {\bibinfo {volume} {25}},\
  \bibinfo {pages} {17} (\bibinfo {year} {2024})}\BibitemShut {NoStop}%
\bibitem [{\citenamefont {Alloul}\ \emph {et~al.}(2009)\citenamefont {Alloul},
  \citenamefont {Bobroff}, \citenamefont {Gabay},\ and\ \citenamefont
  {Hirschfeld}}]{allouletal_RMP}%
  \BibitemOpen
  \bibfield  {author} {\bibinfo {author} {\bibfnamefont {H.}~\bibnamefont
  {Alloul}}, \bibinfo {author} {\bibfnamefont {J.}~\bibnamefont {Bobroff}},
  \bibinfo {author} {\bibfnamefont {M.}~\bibnamefont {Gabay}},\ and\ \bibinfo
  {author} {\bibfnamefont {P.~J.}\ \bibnamefont {Hirschfeld}},\ }\bibfield
  {title} {\bibinfo {title} {Defects in correlated metals and
  superconductors},\ }\href {https://doi.org/10.1103/RevModPhys.81.45}
  {\bibfield  {journal} {\bibinfo  {journal} {Rev. Mod. Phys.}\ }\textbf
  {\bibinfo {volume} {81}},\ \bibinfo {pages} {45} (\bibinfo {year}
  {2009})}\BibitemShut {NoStop}%
\bibitem [{\citenamefont {Yerin}\ and\ \citenamefont
  {Varlamov}(2022)}]{Yerin_multiband_superconductivity_disorder}%
  \BibitemOpen
  \bibfield  {author} {\bibinfo {author} {\bibfnamefont {C.}~\bibnamefont
  {Yerin}, \bibfnamefont {Y.~Petrillo}}\ and\ \bibinfo {author} {\bibfnamefont
  {A.~A.}\ \bibnamefont {Varlamov}},\ }\bibfield  {title} {\bibinfo {title}
  {The lifshitz nature of the transition between the gap and gapless states of
  superconductor},\ }\href {https://doi.org/10.21468/SciPostPhysCore.5.1.009}
  {\bibfield  {journal} {\bibinfo  {journal} {SciPost Phys. Core}\ }\textbf
  {\bibinfo {volume} {5}},\ \bibinfo {pages} {009} (\bibinfo {year}
  {2022})}\BibitemShut {NoStop}%
\bibitem [{\citenamefont {Huscroft}\ and\ \citenamefont
  {Scalettar}(1997)}]{huscroft_hubbard_non_mag_disorder}%
  \BibitemOpen
  \bibfield  {author} {\bibinfo {author} {\bibfnamefont {C.}~\bibnamefont
  {Huscroft}}\ and\ \bibinfo {author} {\bibfnamefont {R.~T.}\ \bibnamefont
  {Scalettar}},\ }\bibfield  {title} {\bibinfo {title} {Effect of disorder on
  charge-density wave and superconducting order in the half-filled attractive
  hubbard model},\ }\href {https://doi.org/10.1103/PhysRevB.55.1185} {\bibfield
   {journal} {\bibinfo  {journal} {Phys. Rev. B}\ }\textbf {\bibinfo {volume}
  {55}},\ \bibinfo {pages} {1185} (\bibinfo {year} {1997})}\BibitemShut
  {NoStop}%
\bibitem [{\citenamefont {Dobrosavljevic}\ and\ \citenamefont
  {Kotliar}(1997)}]{dobrosavljevic_infinited_disorder}%
  \BibitemOpen
  \bibfield  {author} {\bibinfo {author} {\bibfnamefont {V.}~\bibnamefont
  {Dobrosavljevic}}\ and\ \bibinfo {author} {\bibfnamefont {G.}~\bibnamefont
  {Kotliar}},\ }\bibfield  {title} {\bibinfo {title} {Mean field theory of the
  mott-anderson transition},\ }\href
  {https://doi.org/10.1103/PhysRevLett.78.3943} {\bibfield  {journal} {\bibinfo
   {journal} {Physical Review Letters}\ }\textbf {\bibinfo {volume} {78}},\
  \bibinfo {pages} {3943} (\bibinfo {year} {1997})}\BibitemShut {NoStop}%
\bibitem [{\citenamefont {Benfatto}\ \emph {et~al.}(2008)\citenamefont
  {Benfatto}, \citenamefont {Castellani},\ and\ \citenamefont
  {Giamarchi}}]{benfatto_XY_bilayers}%
  \BibitemOpen
  \bibfield  {author} {\bibinfo {author} {\bibfnamefont {L.}~\bibnamefont
  {Benfatto}}, \bibinfo {author} {\bibfnamefont {C.}~\bibnamefont
  {Castellani}},\ and\ \bibinfo {author} {\bibfnamefont {T.}~\bibnamefont
  {Giamarchi}},\ }\bibfield  {title} {\bibinfo {title} {Doping dependence of
  the vortex-core energy in bilayer films of cuprates},\ }\href
  {https://doi.org/10.1103/PhysRevB.77.100506} {\bibfield  {journal} {\bibinfo
  {journal} {Phys. Rev. B}\ }\textbf {\bibinfo {volume} {77}},\ \bibinfo
  {pages} {100506} (\bibinfo {year} {2008})}\BibitemShut {NoStop}%
\bibitem [{\citenamefont {Dubi}\ \emph {et~al.}(2007)\citenamefont {Dubi},
  \citenamefont {Meir},\ and\ \citenamefont {Avishai}}]{dubi_nature07}%
  \BibitemOpen
  \bibfield  {author} {\bibinfo {author} {\bibfnamefont {Y.}~\bibnamefont
  {Dubi}}, \bibinfo {author} {\bibfnamefont {Y.}~\bibnamefont {Meir}},\ and\
  \bibinfo {author} {\bibfnamefont {Y.}~\bibnamefont {Avishai}},\ }\bibfield
  {title} {\bibinfo {title} {Nature of the superconductor--insulator transition
  in disordered superconductors},\ }\href {https://doi.org/10.1038/nature06180}
  {\bibfield  {journal} {\bibinfo  {journal} {Nature}\ }\textbf {\bibinfo
  {volume} {449}},\ \bibinfo {pages} {876} (\bibinfo {year}
  {2007})}\BibitemShut {NoStop}%
\bibitem [{\citenamefont {Anderson}(1958)}]{anderson_localisation}%
  \BibitemOpen
  \bibfield  {author} {\bibinfo {author} {\bibfnamefont {P.~W.}\ \bibnamefont
  {Anderson}},\ }\bibfield  {title} {\bibinfo {title} {Absence of diffusion in
  certain random lattices},\ }\href {https://doi.org/10.1103/PhysRev.109.1492}
  {\bibfield  {journal} {\bibinfo  {journal} {Phys. Rev.}\ }\textbf {\bibinfo
  {volume} {109}},\ \bibinfo {pages} {1492} (\bibinfo {year}
  {1958})}\BibitemShut {NoStop}%
\bibitem [{\citenamefont {Suzumura}\ and\ \citenamefont
  {Giamarchi}(1989)}]{suzumura_mean_field}%
  \BibitemOpen
  \bibfield  {author} {\bibinfo {author} {\bibfnamefont {Y.}~\bibnamefont
  {Suzumura}}\ and\ \bibinfo {author} {\bibfnamefont {T.}~\bibnamefont
  {Giamarchi}},\ }\bibfield  {title} {\bibinfo {title} {Impurity pinning vs
  superconductivity in quasi-one-dimensional electron},\ }\href
  {https://doi.org/https://doi.org/10.1143/JPSJ.58.1748} {\bibfield  {journal}
  {\bibinfo  {journal} {Journal of the Physical Society of Japan}\ }\textbf
  {\bibinfo {volume} {58}},\ \bibinfo {pages} {1748} (\bibinfo {year}
  {1989})}\BibitemShut {NoStop}%
\bibitem [{\citenamefont {Lowe}\ \emph {et~al.}(2021)\citenamefont {Lowe},
  \citenamefont {Kagalovsky},\ and\ \citenamefont
  {Yurkevich}}]{lowe_disorder_enhance_superconductivity_quasi_1d}%
  \BibitemOpen
  \bibfield  {author} {\bibinfo {author} {\bibfnamefont {A.}~\bibnamefont
  {Lowe}}, \bibinfo {author} {\bibfnamefont {V.}~\bibnamefont {Kagalovsky}},\
  and\ \bibinfo {author} {\bibfnamefont {I.~V.}\ \bibnamefont {Yurkevich}},\
  }\bibfield  {title} {\bibinfo {title} {Disorder-enhanced superconductivity in
  a quasi-one-dimensional strongly correlated system},\ }\href
  {https://doi.org/10.1103/PhysRevResearch.3.033059} {\bibfield  {journal}
  {\bibinfo  {journal} {Physical Review Research}\ }\textbf {\bibinfo {volume}
  {3}},\ \bibinfo {pages} {033059} (\bibinfo {year} {2021})}\BibitemShut
  {NoStop}%
\bibitem [{\citenamefont {Berezinskii}(1974)}]{berezinskii_conductivity_log}%
  \BibitemOpen
  \bibfield  {author} {\bibinfo {author} {\bibfnamefont {V.~L.}\ \bibnamefont
  {Berezinskii}},\ }\bibfield  {title} {\bibinfo {title} {Kinetics of a quantum
  particle in a one-dimensional random potential},\ }\href
  {http://jetp.ras.ru/cgi-bin/dn/e_038_03_0620.pdf} {\bibfield  {journal}
  {\bibinfo  {journal} {Journal of Experimental and Theoretical Physics}\
  }\textbf {\bibinfo {volume} {38}},\ \bibinfo {pages} {620} (\bibinfo {year}
  {1974})}\BibitemShut {NoStop}%
\bibitem [{\citenamefont {Abrikosov}\ and\ \citenamefont
  {Rhyzkin}(1978)}]{abrikosov_rhyzkin}%
  \BibitemOpen
  \bibfield  {author} {\bibinfo {author} {\bibfnamefont {A.~A.}\ \bibnamefont
  {Abrikosov}}\ and\ \bibinfo {author} {\bibfnamefont {J.~A.}\ \bibnamefont
  {Rhyzkin}},\ }\bibfield  {title} {\bibinfo {title} {Conductivity of
  quasi-one-dimensional metal systems},\ }\href
  {https://doi.org/10.1080/00018737800101364} {\bibfield  {journal} {\bibinfo
  {journal} {adv. Phys.}\ }\textbf {\bibinfo {volume} {27}},\ \bibinfo {pages}
  {147} (\bibinfo {year} {1978})}\BibitemShut {NoStop}%
\bibitem [{\citenamefont {Giamarchi}\ and\ \citenamefont
  {Schulz}(1988)}]{giamarchi_loc}%
  \BibitemOpen
  \bibfield  {author} {\bibinfo {author} {\bibfnamefont {T.}~\bibnamefont
  {Giamarchi}}\ and\ \bibinfo {author} {\bibfnamefont {H.~J.}\ \bibnamefont
  {Schulz}},\ }\bibfield  {title} {\bibinfo {title} {Anderson localization and
  interactions in one-dimensional metals},\ }\href
  {https://doi.org/10.1103/PhysRevB.37.325} {\bibfield  {journal} {\bibinfo
  {journal} {Phys. Rev. B}\ }\textbf {\bibinfo {volume} {37}},\ \bibinfo
  {pages} {325} (\bibinfo {year} {1988})}\BibitemShut {NoStop}%
\bibitem [{\citenamefont {Bloch}\ \emph {et~al.}(2008)\citenamefont {Bloch},
  \citenamefont {Dalibard},\ and\ \citenamefont {Zwerger}}]{Bloch2008}%
  \BibitemOpen
  \bibfield  {author} {\bibinfo {author} {\bibfnamefont {I.}~\bibnamefont
  {Bloch}}, \bibinfo {author} {\bibfnamefont {J.}~\bibnamefont {Dalibard}},\
  and\ \bibinfo {author} {\bibfnamefont {W.}~\bibnamefont {Zwerger}},\
  }\bibfield  {title} {\bibinfo {title} {Many-body physics with ultracold
  gases},\ }\href {https://doi.org/10.1103/RevModPhys.80.885} {\bibfield
  {journal} {\bibinfo  {journal} {Rev. Mod. Phys.}\ }\textbf {\bibinfo {volume}
  {80}},\ \bibinfo {pages} {885} (\bibinfo {year} {2008})}\BibitemShut
  {NoStop}%
\bibitem [{\citenamefont {{Esslinger}}(2010)}]{esslinger_annrev_2010}%
  \BibitemOpen
  \bibfield  {author} {\bibinfo {author} {\bibfnamefont {T.}~\bibnamefont
  {{Esslinger}}},\ }\bibfield  {title} {\bibinfo {title} {{Fermi-Hubbard
  Physics with Atoms in an Optical Lattice}},\ }\href
  {https://doi.org/10.1146/annurev-conmatphys-070909-104059} {\bibfield
  {journal} {\bibinfo  {journal} {Annual Review of Condensed Matter Physics}\
  }\textbf {\bibinfo {volume} {1}},\ \bibinfo {pages} {129} (\bibinfo {year}
  {2010})},\ \Eprint {https://arxiv.org/abs/1007.0012} {1007.0012} \BibitemShut
  {NoStop}%
\bibitem [{\citenamefont {Roati}\ \emph {et~al.}(2008)\citenamefont {Roati},
  \citenamefont {D'Errico}, \citenamefont {Fallani}, \citenamefont {Fattori},
  \citenamefont {Fort}, \citenamefont {Zaccanti}, \citenamefont {Modugno},
  \citenamefont {Modugno},\ and\ \citenamefont
  {Inguscio}}]{roati08_anderson_localization_BEC}%
  \BibitemOpen
  \bibfield  {author} {\bibinfo {author} {\bibfnamefont {G.}~\bibnamefont
  {Roati}}, \bibinfo {author} {\bibfnamefont {C.}~\bibnamefont {D'Errico}},
  \bibinfo {author} {\bibfnamefont {L.}~\bibnamefont {Fallani}}, \bibinfo
  {author} {\bibfnamefont {M.}~\bibnamefont {Fattori}}, \bibinfo {author}
  {\bibfnamefont {C.}~\bibnamefont {Fort}}, \bibinfo {author} {\bibfnamefont
  {M.}~\bibnamefont {Zaccanti}}, \bibinfo {author} {\bibfnamefont
  {G.}~\bibnamefont {Modugno}}, \bibinfo {author} {\bibfnamefont
  {M.}~\bibnamefont {Modugno}},\ and\ \bibinfo {author} {\bibfnamefont
  {M.}~\bibnamefont {Inguscio}},\ }\bibfield  {title} {\bibinfo {title}
  {Anderson localization of a non-interacting bose-einstein condensate},\
  }\href {https://doi.org/10.1038/nature07071} {\bibfield  {journal} {\bibinfo
  {journal} {Nature}\ }\textbf {\bibinfo {volume} {453}},\ \bibinfo {pages}
  {895} (\bibinfo {year} {2008})}\BibitemShut {NoStop}%
\bibitem [{\citenamefont {Billy}\ \emph {et~al.}(2008)\citenamefont {Billy},
  \citenamefont {Josse}, \citenamefont {Zuo}, \citenamefont {Bernard},
  \citenamefont {Hambrecht}, \citenamefont {Lugan}, \citenamefont {Clement},
  \citenamefont {Sanchez-Palencia}, \citenamefont {Bouyer},\ and\ \citenamefont
  {Aspect}}]{aspect_anderson_localization_BEC.pdf}%
  \BibitemOpen
  \bibfield  {author} {\bibinfo {author} {\bibfnamefont {J.}~\bibnamefont
  {Billy}}, \bibinfo {author} {\bibfnamefont {V.}~\bibnamefont {Josse}},
  \bibinfo {author} {\bibfnamefont {Z.}~\bibnamefont {Zuo}}, \bibinfo {author}
  {\bibfnamefont {A.}~\bibnamefont {Bernard}}, \bibinfo {author} {\bibfnamefont
  {B.}~\bibnamefont {Hambrecht}}, \bibinfo {author} {\bibfnamefont
  {P.}~\bibnamefont {Lugan}}, \bibinfo {author} {\bibfnamefont
  {D.}~\bibnamefont {Clement}}, \bibinfo {author} {\bibfnamefont
  {L.}~\bibnamefont {Sanchez-Palencia}}, \bibinfo {author} {\bibfnamefont
  {P.}~\bibnamefont {Bouyer}},\ and\ \bibinfo {author} {\bibfnamefont
  {A.}~\bibnamefont {Aspect}},\ }\bibfield  {title} {\bibinfo {title} {Direct
  observation of anderson localization of matter waves in a controlled
  disorder},\ }\href {https://doi.org/10.1038/nature07000} {\bibfield
  {journal} {\bibinfo  {journal} {Nature}\ }\textbf {\bibinfo {volume} {453}},\
  \bibinfo {pages} {891} (\bibinfo {year} {2008})}\BibitemShut {NoStop}%
\bibitem [{\citenamefont {Sanchez-Palencia}\ and\ \citenamefont
  {Lewenstein}(2010)}]{sanchez-palencia_review_disorder_cold}%
  \BibitemOpen
  \bibfield  {author} {\bibinfo {author} {\bibfnamefont {L.}~\bibnamefont
  {Sanchez-Palencia}}\ and\ \bibinfo {author} {\bibfnamefont {M.}~\bibnamefont
  {Lewenstein}},\ }\bibfield  {title} {\bibinfo {title} {Disordered quantum
  gases under control},\ }\href {https://doi.org/10.1038/nphys1507} {\bibfield
  {journal} {\bibinfo  {journal} {Nat. Phys.}\ }\textbf {\bibinfo {volume}
  {6}},\ \bibinfo {pages} {87} (\bibinfo {year} {2010})}\BibitemShut {NoStop}%
\bibitem [{\citenamefont {Piraud}\ and\ \citenamefont
  {Sanchez-Palencia}(2013)}]{Piraud_Anderson_speckle}%
  \BibitemOpen
  \bibfield  {author} {\bibinfo {author} {\bibfnamefont {M.}~\bibnamefont
  {Piraud}}\ and\ \bibinfo {author} {\bibfnamefont {L.}~\bibnamefont
  {Sanchez-Palencia}},\ }\bibfield  {title} {\bibinfo {title} {Tailoring
  anderson localization by disorder correlations in 1d speckle potentials},\
  }\href {https://doi.org/10.1140/epjst/e2013-01758-6} {\bibfield  {journal}
  {\bibinfo  {journal} {Eur. Phys. J. Special Topics}\ }\textbf {\bibinfo
  {volume} {217}},\ \bibinfo {pages} {91} (\bibinfo {year} {2013})}\BibitemShut
  {NoStop}%
\bibitem [{\citenamefont {Fallani}\ \emph {et~al.}(2007)\citenamefont
  {Fallani}, \citenamefont {Lye}, \citenamefont {Guarrera}, \citenamefont
  {Fort},\ and\ \citenamefont {Inguscio}}]{fallani_biperiodic_cold}%
  \BibitemOpen
  \bibfield  {author} {\bibinfo {author} {\bibfnamefont {L.}~\bibnamefont
  {Fallani}}, \bibinfo {author} {\bibfnamefont {J.~E.}\ \bibnamefont {Lye}},
  \bibinfo {author} {\bibfnamefont {V.}~\bibnamefont {Guarrera}}, \bibinfo
  {author} {\bibfnamefont {C.}~\bibnamefont {Fort}},\ and\ \bibinfo {author}
  {\bibfnamefont {M.}~\bibnamefont {Inguscio}},\ }\bibfield  {title} {\bibinfo
  {title} {Ultracold atoms in a disordered crystal of light: Towards a bose
  glass},\ }\href {https://doi.org/10.1103/PhysRevLett.98.130404} {\bibfield
  {journal} {\bibinfo  {journal} {Physical Review Letters}\ }\textbf {\bibinfo
  {volume} {98}},\ \bibinfo {pages} {130404} (\bibinfo {year}
  {2007})}\BibitemShut {NoStop}%
\bibitem [{\citenamefont {D'Errico}\ \emph {et~al.}(2014)\citenamefont
  {D'Errico}, \citenamefont {Lucioni}, \citenamefont {Tanzi}, \citenamefont
  {Gori}, \citenamefont {Roux}, \citenamefont {McCulloch}, \citenamefont
  {Giamarchi}, \citenamefont {Inguscio},\ and\ \citenamefont
  {Modugno}}]{derrico_cold_boseglass}%
  \BibitemOpen
  \bibfield  {author} {\bibinfo {author} {\bibfnamefont {C.}~\bibnamefont
  {D'Errico}}, \bibinfo {author} {\bibfnamefont {E.}~\bibnamefont {Lucioni}},
  \bibinfo {author} {\bibfnamefont {L.}~\bibnamefont {Tanzi}}, \bibinfo
  {author} {\bibfnamefont {L.}~\bibnamefont {Gori}}, \bibinfo {author}
  {\bibfnamefont {G.}~\bibnamefont {Roux}}, \bibinfo {author} {\bibfnamefont
  {I.~P.}\ \bibnamefont {McCulloch}}, \bibinfo {author} {\bibfnamefont
  {T.}~\bibnamefont {Giamarchi}}, \bibinfo {author} {\bibfnamefont
  {M.}~\bibnamefont {Inguscio}},\ and\ \bibinfo {author} {\bibfnamefont
  {G.}~\bibnamefont {Modugno}},\ }\bibfield  {title} {\bibinfo {title}
  {Observation of a disordered bosonic insulator from weak to strong
  interactions},\ }\href {https://doi.org/10.1103/PhysRevLett.113.095301}
  {\bibfield  {journal} {\bibinfo  {journal} {Phys. Rev. Lett.}\ }\textbf
  {\bibinfo {volume} {113}},\ \bibinfo {pages} {095301} (\bibinfo {year}
  {2014})}\BibitemShut {NoStop}%
\bibitem [{\citenamefont {Jendrzejeweski}\ \emph {et~al.}(2012)\citenamefont
  {Jendrzejeweski}, \citenamefont {Bernard}, \citenamefont {Müller},
  \citenamefont {Cheinet}, \citenamefont {Josse}, \citenamefont {Piraud},
  \citenamefont {Pezzé}, \citenamefont {Sanchez-Palencia}, \citenamefont
  {Aspect},\ and\ \citenamefont
  {Bouyer}}]{jendrzejewski_3d_anderson_localisazion_cold_atoms}%
  \BibitemOpen
  \bibfield  {author} {\bibinfo {author} {\bibfnamefont {F.}~\bibnamefont
  {Jendrzejeweski}}, \bibinfo {author} {\bibfnamefont {A.}~\bibnamefont
  {Bernard}}, \bibinfo {author} {\bibfnamefont {K.}~\bibnamefont {Müller}},
  \bibinfo {author} {\bibfnamefont {P.}~\bibnamefont {Cheinet}}, \bibinfo
  {author} {\bibfnamefont {V.}~\bibnamefont {Josse}}, \bibinfo {author}
  {\bibfnamefont {M.}~\bibnamefont {Piraud}}, \bibinfo {author} {\bibfnamefont
  {L.}~\bibnamefont {Pezzé}}, \bibinfo {author} {\bibfnamefont
  {L.}~\bibnamefont {Sanchez-Palencia}}, \bibinfo {author} {\bibfnamefont
  {A.}~\bibnamefont {Aspect}},\ and\ \bibinfo {author} {\bibfnamefont
  {P.}~\bibnamefont {Bouyer}},\ }\bibfield  {title} {\bibinfo {title}
  {Three-dimensional localization of ultracold atoms in an optical disordered
  potential},\ }\href {https://doi.org/10.1038/nphys2256} {\bibfield  {journal}
  {\bibinfo  {journal} {Nature Physics}\ }\textbf {\bibinfo {volume} {8}},\
  \bibinfo {pages} {398} (\bibinfo {year} {2012})}\BibitemShut {NoStop}%
\bibitem [{\citenamefont {Schreiber}\ \emph {et~al.}(2015)\citenamefont
  {Schreiber}, \citenamefont {Hodgman}, \citenamefont {Bordia}, \citenamefont
  {L{\"u}schen}, \citenamefont {Fischer}, \citenamefont {Vosk}, \citenamefont
  {Altman}, \citenamefont {Schneider},\ and\ \citenamefont
  {Bloch}}]{schreiber_manybody_localization_cold}%
  \BibitemOpen
  \bibfield  {author} {\bibinfo {author} {\bibfnamefont {M.}~\bibnamefont
  {Schreiber}}, \bibinfo {author} {\bibfnamefont {S.~S.}\ \bibnamefont
  {Hodgman}}, \bibinfo {author} {\bibfnamefont {P.}~\bibnamefont {Bordia}},
  \bibinfo {author} {\bibfnamefont {H.~P.}\ \bibnamefont {L{\"u}schen}},
  \bibinfo {author} {\bibfnamefont {M.~H.}\ \bibnamefont {Fischer}}, \bibinfo
  {author} {\bibfnamefont {R.}~\bibnamefont {Vosk}}, \bibinfo {author}
  {\bibfnamefont {E.}~\bibnamefont {Altman}}, \bibinfo {author} {\bibfnamefont
  {U.}~\bibnamefont {Schneider}},\ and\ \bibinfo {author} {\bibfnamefont
  {I.}~\bibnamefont {Bloch}},\ }\bibfield  {title} {\bibinfo {title}
  {Observation of many-body localization of interacting fermions in a
  quasirandom optical lattice},\ }\href
  {https://doi.org/10.1126/science.aaa7432} {\bibfield  {journal} {\bibinfo
  {journal} {Science}\ }\textbf {\bibinfo {volume} {349}},\ \bibinfo {pages}
  {842} (\bibinfo {year} {2015})}\BibitemShut {NoStop}%
\bibitem [{\citenamefont {Yao}\ \emph {et~al.}(2019)\citenamefont {Yao},
  \citenamefont {Khoudli}, \citenamefont {Bresque},\ and\ \citenamefont
  {Sanchez-Palencia}}]{yao_quasiperiodic_1d_equal_potential}%
  \BibitemOpen
  \bibfield  {author} {\bibinfo {author} {\bibfnamefont {H.}~\bibnamefont
  {Yao}}, \bibinfo {author} {\bibfnamefont {A.}~\bibnamefont {Khoudli}},
  \bibinfo {author} {\bibfnamefont {L.}~\bibnamefont {Bresque}},\ and\ \bibinfo
  {author} {\bibfnamefont {L.}~\bibnamefont {Sanchez-Palencia}},\ }\bibfield
  {title} {\bibinfo {title} {Critical behavior and fractality in shallow
  one-dimensional quasiperiodic potentials},\ }\href
  {https://doi.org/10.1103/PhysRevLett.123.070405} {\bibfield  {journal}
  {\bibinfo  {journal} {Physical Review Letters}\ }\textbf {\bibinfo {volume}
  {123}},\ \bibinfo {pages} {070405} (\bibinfo {year} {2019})}\BibitemShut
  {NoStop}%
\bibitem [{\citenamefont {Yao}\ \emph {et~al.}(2020)\citenamefont {Yao},
  \citenamefont {Giamarchi},\ and\ \citenamefont
  {Sanchez-Palencia}}]{yao_quasiperiodic_equal_potential_bose_glass}%
  \BibitemOpen
  \bibfield  {author} {\bibinfo {author} {\bibfnamefont {H.}~\bibnamefont
  {Yao}}, \bibinfo {author} {\bibfnamefont {T.}~\bibnamefont {Giamarchi}},\
  and\ \bibinfo {author} {\bibfnamefont {L.}~\bibnamefont {Sanchez-Palencia}},\
  }\bibfield  {title} {\bibinfo {title} {Lieb-liniger bosons in a shallow
  quasiperiodic potential: Bose glass phase and fractal mott lobes},\ }\href
  {https://doi.org/10.1103/PhysRevLett.125.060401} {\bibfield  {journal}
  {\bibinfo  {journal} {Physical Review Letters}\ }\textbf {\bibinfo {volume}
  {125}},\ \bibinfo {pages} {060401} (\bibinfo {year} {2020})}\BibitemShut
  {NoStop}%
\bibitem [{\citenamefont {Sbroscia}\ \emph {et~al.}(2020)\citenamefont
  {Sbroscia}, \citenamefont {Viebahn}, \citenamefont {Carter}, \citenamefont
  {Yu}, \citenamefont {Gaunt},\ and\ \citenamefont
  {Schneider}}]{Sbroscia_2d_quasicrstal_localization}%
  \BibitemOpen
  \bibfield  {author} {\bibinfo {author} {\bibfnamefont {M.}~\bibnamefont
  {Sbroscia}}, \bibinfo {author} {\bibfnamefont {K.}~\bibnamefont {Viebahn}},
  \bibinfo {author} {\bibfnamefont {E.}~\bibnamefont {Carter}}, \bibinfo
  {author} {\bibfnamefont {J.-C.}\ \bibnamefont {Yu}}, \bibinfo {author}
  {\bibfnamefont {A.}~\bibnamefont {Gaunt}},\ and\ \bibinfo {author}
  {\bibfnamefont {U.}~\bibnamefont {Schneider}},\ }\bibfield  {title} {\bibinfo
  {title} {Observing localization in a 2d quasicrytsalline optical lattice},\
  }\href {https://doi.org/10.1103/PhysRevLett.125.200604} {\bibfield  {journal}
  {\bibinfo  {journal} {Physical Review Letters}\ }\textbf {\bibinfo {volume}
  {125}},\ \bibinfo {pages} {200604} (\bibinfo {year} {2020})}\BibitemShut
  {NoStop}%
\bibitem [{\citenamefont {Giamarchi}(2004)}]{giamarchi_book_1d}%
  \BibitemOpen
  \bibfield  {author} {\bibinfo {author} {\bibfnamefont {T.}~\bibnamefont
  {Giamarchi}},\ }\href@noop {} {\emph {\bibinfo {title} {Quantum Physics in
  One Dimension}}},\ \bibinfo {series} {International series of monographs on
  physics}, Vol.\ \bibinfo {volume} {121}\ (\bibinfo  {publisher} {Oxford
  University Press},\ \bibinfo {address} {Oxford},\ \bibinfo {year}
  {2004})\BibitemShut {NoStop}%
\bibitem [{\citenamefont {Lugan}\ \emph {et~al.}(2009)\citenamefont {Lugan},
  \citenamefont {Aspect}, \citenamefont {Sanchez-Palencia}, \citenamefont
  {Delande}, \citenamefont {Gr\'emaud}, \citenamefont {M\"uller},\ and\
  \citenamefont {Miniatura}}]{lugan_correlated_potentials}%
  \BibitemOpen
  \bibfield  {author} {\bibinfo {author} {\bibfnamefont {P.}~\bibnamefont
  {Lugan}}, \bibinfo {author} {\bibfnamefont {A.}~\bibnamefont {Aspect}},
  \bibinfo {author} {\bibfnamefont {L.}~\bibnamefont {Sanchez-Palencia}},
  \bibinfo {author} {\bibfnamefont {D.}~\bibnamefont {Delande}}, \bibinfo
  {author} {\bibfnamefont {B.}~\bibnamefont {Gr\'emaud}}, \bibinfo {author}
  {\bibfnamefont {C.~A.}\ \bibnamefont {M\"uller}},\ and\ \bibinfo {author}
  {\bibfnamefont {C.}~\bibnamefont {Miniatura}},\ }\bibfield  {title} {\bibinfo
  {title} {One-dimensional anderson localization in certain correlated random
  potentials},\ }\href {https://doi.org/10.1103/PhysRevA.80.023605} {\bibfield
  {journal} {\bibinfo  {journal} {Phys. Rev. A}\ }\textbf {\bibinfo {volume}
  {80}},\ \bibinfo {pages} {023605} (\bibinfo {year} {2009})}\BibitemShut
  {NoStop}%
\bibitem [{\citenamefont {Bourbonnais}\ and\ \citenamefont
  {Caron}(1988)}]{bourbonnais_lettre_tperp_RG}%
  \BibitemOpen
  \bibfield  {author} {\bibinfo {author} {\bibfnamefont {C.}~\bibnamefont
  {Bourbonnais}}\ and\ \bibinfo {author} {\bibfnamefont {L.~G.}\ \bibnamefont
  {Caron}},\ }\bibfield  {title} {\bibinfo {title} {New mechanisms for phase
  transitions in quasi-one-dimensional conductors},\ }\href
  {https://doi.org/10.1209/0295-5075/5/3/005} {\bibfield  {journal} {\bibinfo
  {journal} {Europhysics Letters}\ }\textbf {\bibinfo {volume} {5}},\ \bibinfo
  {pages} {209} (\bibinfo {year} {1988})}\BibitemShut {NoStop}%
\bibitem [{\citenamefont {Giamarchi}(1992)}]{giamarchi_attract_1d}%
  \BibitemOpen
  \bibfield  {author} {\bibinfo {author} {\bibfnamefont {T.}~\bibnamefont
  {Giamarchi}},\ }\bibfield  {title} {\bibinfo {title} {Resistivity of a
  one-dimensional interacting quantum fluid},\ }\href
  {https://doi.org/10.1103/PhysRevB.46.342} {\bibfield  {journal} {\bibinfo
  {journal} {Physical Review B}\ }\textbf {\bibinfo {volume} {46}},\ \bibinfo
  {pages} {342} (\bibinfo {year} {1992})}\BibitemShut {NoStop}%
\bibitem [{\citenamefont {Orignac}\ \emph {et~al.}(1999)\citenamefont
  {Orignac}, \citenamefont {Giamarchi},\ and\ \citenamefont
  {Doussal}}]{orignac_mg_short}%
  \BibitemOpen
  \bibfield  {author} {\bibinfo {author} {\bibfnamefont {E.}~\bibnamefont
  {Orignac}}, \bibinfo {author} {\bibfnamefont {T.}~\bibnamefont {Giamarchi}},\
  and\ \bibinfo {author} {\bibfnamefont {P.~L.}\ \bibnamefont {Doussal}},\
  }\bibfield  {title} {\bibinfo {title} {Possible new phase of commensurate
  insulators with disorder: The mott glass},\ }\href
  {https://doi.org/10.1103/PhysRevLett.83.2378} {\bibfield  {journal} {\bibinfo
   {journal} {Physical Review Letters}\ }\textbf {\bibinfo {volume} {83}},\
  \bibinfo {pages} {2378} (\bibinfo {year} {1999})}\BibitemShut {NoStop}%
\bibitem [{\citenamefont {Giamarchi}\ \emph {et~al.}(2001)\citenamefont
  {Giamarchi}, \citenamefont {{Le Doussal}},\ and\ \citenamefont
  {Orignac}}]{giamarchi_mottglass_long}%
  \BibitemOpen
  \bibfield  {author} {\bibinfo {author} {\bibfnamefont {T.}~\bibnamefont
  {Giamarchi}}, \bibinfo {author} {\bibfnamefont {P.}~\bibnamefont {{Le
  Doussal}}},\ and\ \bibinfo {author} {\bibfnamefont {E.}~\bibnamefont
  {Orignac}},\ }\bibfield  {title} {\bibinfo {title} {Competition of random and
  periodic potentials in interacting fermionic systems and classical
  equivalents:the mott glass},\ }\href
  {https://doi.org/10.1103/PhysRevB.64.245119} {\bibfield  {journal} {\bibinfo
  {journal} {Physical Review B}\ }\textbf {\bibinfo {volume} {64}},\ \bibinfo
  {pages} {245119} (\bibinfo {year} {2001})}\BibitemShut {NoStop}%
\bibitem [{\citenamefont {Giamarchi}\ and\ \citenamefont
  {Schulz}(1989)}]{giamarchi_logs}%
  \BibitemOpen
  \bibfield  {author} {\bibinfo {author} {\bibfnamefont {T.}~\bibnamefont
  {Giamarchi}}\ and\ \bibinfo {author} {\bibfnamefont {H.~J.}\ \bibnamefont
  {Schulz}},\ }\bibfield  {title} {\bibinfo {title} {Correlation functions of
  one-dimensional quantum systems},\ }\href
  {https://doi.org/10.1103/PhysRevB.39.4620} {\bibfield  {journal} {\bibinfo
  {journal} {Physical Review B}\ }\textbf {\bibinfo {volume} {39}},\ \bibinfo
  {pages} {4620} (\bibinfo {year} {1989})}\BibitemShut {NoStop}%
\bibitem [{\citenamefont {Horowitz}\ \emph {et~al.}(1983)\citenamefont
  {Horowitz}, \citenamefont {Bohr}, \citenamefont {Kosterlitz},\ and\
  \citenamefont {Schulz}}]{horowitz_renormalization_incommensurable}%
  \BibitemOpen
  \bibfield  {author} {\bibinfo {author} {\bibfnamefont {B.}~\bibnamefont
  {Horowitz}}, \bibinfo {author} {\bibfnamefont {T.}~\bibnamefont {Bohr}},
  \bibinfo {author} {\bibfnamefont {J.~M.}\ \bibnamefont {Kosterlitz}},\ and\
  \bibinfo {author} {\bibfnamefont {H.~J.}\ \bibnamefont {Schulz}},\ }\bibfield
   {title} {\bibinfo {title} {Commensurate-incommensurate transitions and a
  floating devil’s staircase},\ }\href
  {https://doi.org/10.1103/PhysRevB.28.6596} {\bibfield  {journal} {\bibinfo
  {journal} {Physical Review B}\ }\textbf {\bibinfo {volume} {28}},\ \bibinfo
  {pages} {6596} (\bibinfo {year} {1983})}\BibitemShut {NoStop}%
\bibitem [{\citenamefont {Hart}\ \emph {et~al.}(2015)\citenamefont {Hart},
  \citenamefont {Duarte}, \citenamefont {Yang}, \citenamefont {Liu},
  \citenamefont {Paiva}, \citenamefont {Khatami}, \citenamefont {Scalettar},
  \citenamefont {Trivedi}, \citenamefont {Huse},\ and\ \citenamefont
  {Hulet}}]{hart_antiferro_Hubbard_ultracold_atoms}%
  \BibitemOpen
  \bibfield  {author} {\bibinfo {author} {\bibfnamefont {R.~A.}\ \bibnamefont
  {Hart}}, \bibinfo {author} {\bibfnamefont {P.~M.}\ \bibnamefont {Duarte}},
  \bibinfo {author} {\bibfnamefont {T.}~\bibnamefont {Yang}}, \bibinfo {author}
  {\bibfnamefont {X.}~\bibnamefont {Liu}}, \bibinfo {author} {\bibfnamefont
  {T.}~\bibnamefont {Paiva}}, \bibinfo {author} {\bibfnamefont
  {E.}~\bibnamefont {Khatami}}, \bibinfo {author} {\bibfnamefont {R.~T.}\
  \bibnamefont {Scalettar}}, \bibinfo {author} {\bibfnamefont {N.}~\bibnamefont
  {Trivedi}}, \bibinfo {author} {\bibfnamefont {D.~A.}\ \bibnamefont {Huse}},\
  and\ \bibinfo {author} {\bibfnamefont {R.~G.}\ \bibnamefont {Hulet}},\
  }\bibfield  {title} {\bibinfo {title} {Observation of antiferromagnetic
  correlations in the hubbard model with ultracold atoms},\ }\href
  {https://doi.org/10.1038/nature14223} {\bibfield  {journal} {\bibinfo
  {journal} {Nature}\ }\textbf {\bibinfo {volume} {519}},\ \bibinfo {pages}
  {211} (\bibinfo {year} {2015})}\BibitemShut {NoStop}%
\bibitem [{\citenamefont {Boll}\ \emph {et~al.}(2016)\citenamefont {Boll},
  \citenamefont {Hilker}, \citenamefont {Salomon}, \citenamefont {Omran},
  \citenamefont {Nespolo}, \citenamefont {Pollet}, \citenamefont {Bloch},\ and\
  \citenamefont {Gross}}]{boll_quantum_microscope_spin_resolved}%
  \BibitemOpen
  \bibfield  {author} {\bibinfo {author} {\bibfnamefont {M.}~\bibnamefont
  {Boll}}, \bibinfo {author} {\bibfnamefont {T.~A.}\ \bibnamefont {Hilker}},
  \bibinfo {author} {\bibfnamefont {G.}~\bibnamefont {Salomon}}, \bibinfo
  {author} {\bibfnamefont {A.}~\bibnamefont {Omran}}, \bibinfo {author}
  {\bibfnamefont {J.}~\bibnamefont {Nespolo}}, \bibinfo {author} {\bibfnamefont
  {L.}~\bibnamefont {Pollet}}, \bibinfo {author} {\bibfnamefont
  {I.}~\bibnamefont {Bloch}},\ and\ \bibinfo {author} {\bibfnamefont
  {C.}~\bibnamefont {Gross}},\ }\bibfield  {title} {\bibinfo {title} {Spin- and
  density resolved microscopy of antiferro-magnetic correlations in
  fermi-hubbard chains},\ }\href {https://doi.org/10.1126/science.aag1635}
  {\bibfield  {journal} {\bibinfo  {journal} {Science}\ }\textbf {\bibinfo
  {volume} {353}},\ \bibinfo {pages} {1257} (\bibinfo {year}
  {2016})}\BibitemShut {NoStop}%
\bibitem [{\citenamefont {Bakr}\ \emph {et~al.}(2009)\citenamefont {Bakr},
  \citenamefont {Gillen}, \citenamefont {Peng}, \citenamefont {F\"olling},\
  and\ \citenamefont {Greiner}}]{bakr_gillen_09}%
  \BibitemOpen
  \bibfield  {author} {\bibinfo {author} {\bibfnamefont {W.~S.}\ \bibnamefont
  {Bakr}}, \bibinfo {author} {\bibfnamefont {J.~I.}\ \bibnamefont {Gillen}},
  \bibinfo {author} {\bibfnamefont {A.}~\bibnamefont {Peng}}, \bibinfo {author}
  {\bibfnamefont {S.}~\bibnamefont {F\"olling}},\ and\ \bibinfo {author}
  {\bibfnamefont {M.}~\bibnamefont {Greiner}},\ }\bibfield  {title} {\bibinfo
  {title} {A quantum gas microscope for detecting single atoms in a
  hubbard-regime optical lattice},\ }\href
  {https://doi.org/10.1038/nature08482} {\bibfield  {journal} {\bibinfo
  {journal} {Nature}\ }\textbf {\bibinfo {volume} {462}},\ \bibinfo {pages}
  {74} (\bibinfo {year} {2009})}\BibitemShut {NoStop}%
\bibitem [{\citenamefont {Ho}\ \emph {et~al.}(2009)\citenamefont {Ho},
  \citenamefont {Cazalilla},\ and\ \citenamefont
  {Giamarchi}}]{ho_attractive_hubbard}%
  \BibitemOpen
  \bibfield  {author} {\bibinfo {author} {\bibfnamefont {A.~F.}\ \bibnamefont
  {Ho}}, \bibinfo {author} {\bibfnamefont {M.~A.}\ \bibnamefont {Cazalilla}},\
  and\ \bibinfo {author} {\bibfnamefont {T.}~\bibnamefont {Giamarchi}},\
  }\bibfield  {title} {\bibinfo {title} {Quantum simulation of the hubbard
  model: The attractive route},\ }\href
  {https://doi.org/10.1103/PhysRevA.79.033620} {\bibfield  {journal} {\bibinfo
  {journal} {Phys. Rev. A}\ }\textbf {\bibinfo {volume} {79}},\ \bibinfo
  {pages} {033620} (\bibinfo {year} {2009})}\BibitemShut {NoStop}%
\bibitem [{\citenamefont {Wu}\ \emph {et~al.}(2022)\citenamefont {Wu},
  \citenamefont {Gutierrez-Lezama}, \citenamefont {Lopez-Paz}, \citenamefont
  {Gibertini}, \citenamefont {Watanabe}, \citenamefont {Taniguchi},
  \citenamefont {von Rohr}, \citenamefont {Ubrig},\ and\ \citenamefont
  {Morpurgo}}]{fan_CrSBr}%
  \BibitemOpen
  \bibfield  {author} {\bibinfo {author} {\bibfnamefont {F.}~\bibnamefont
  {Wu}}, \bibinfo {author} {\bibfnamefont {I.}~\bibnamefont
  {Gutierrez-Lezama}}, \bibinfo {author} {\bibfnamefont {S.~A.}\ \bibnamefont
  {Lopez-Paz}}, \bibinfo {author} {\bibfnamefont {M.}~\bibnamefont
  {Gibertini}}, \bibinfo {author} {\bibfnamefont {K.}~\bibnamefont {Watanabe}},
  \bibinfo {author} {\bibfnamefont {T.}~\bibnamefont {Taniguchi}}, \bibinfo
  {author} {\bibfnamefont {F.~O.}\ \bibnamefont {von Rohr}}, \bibinfo {author}
  {\bibfnamefont {N.}~\bibnamefont {Ubrig}},\ and\ \bibinfo {author}
  {\bibfnamefont {A.~F.}\ \bibnamefont {Morpurgo}},\ }\bibfield  {title}
  {\bibinfo {title} {Quasi-1d electronic transport in a 2d magnetic
  semiconductor},\ }\href {https://doi.org/10.1002/adma.202109759} {\bibfield
  {journal} {\bibinfo  {journal} {Advanced Materials}\ }\textbf {\bibinfo
  {volume} {34}},\ \bibinfo {pages} {2109759} (\bibinfo {year}
  {2022})}\BibitemShut {NoStop}%
\end{thebibliography}
%

\end{document}